# From traffic conflict simulation to traffic crash simulation: introducing traffic safety indicators based on the explicit simulation of potential driver errors

Vittorio Astarita[*] and Vincenzo Pasquale Giofré

*Università della Calabria, Rende (CS) 87040, Italy*

---

[*] Corresponding author. Tel.: +39-0984-49-6780; fax: +39-0984-49-6787.
*E-mail address*: vittorio.astarita@unical.it

# From traffic conflict simulation to traffic crash simulation: introducing traffic safety indicators based on the explicit simulation of potential driver errors

Vittorio Astarita[*] and Vincenzo Pasquale Giofré

*Università della Calabria, Rende (CS) 87040, Italy*

**Abstract**

Traffic safety conflict indicators have been used to assess the level of safety in different traffic conditions. Traffic safety indicators can be evaluated by numerical elaboration given vehicles trajectories. Real vehicle trajectories are not always easy to obtain and often microsimulation has been used to generate vehicular trajectories on which to apply numerical elaboration to evaluate surrogate safety measures for different scenarios.

However, commonly used traffic conflict indicators do not consider the severity of a potential resulting crash. Only recent works propose to correct common indicators to take into account the amount of energy involved in near-crashes events. The use of conflict-energy-based indicators in simulation has not been yet well explored and could offer interesting results promoting the use of microsimulation for road safety evaluations.

Moreover, commonly used traffic conflict indicators do not consider roadside obstacles or barriers and vehicles which are travelling on non-conflicting trajectories.

This paper presents a complete new approach for microscopic evaluation of traffic safety. The proposed safety indicators are based on the explicit simulation of possible driver errors and possible crashes.

Drivers can be distracted in many different ways such as by being occupied in some other tasks (for example a mobile call) or by a momentary breakdown in attention and/or awareness to external changes due to some psychological or physical failure.

This paper originated from the following consideration: in the real world, most accidents involving human drivers occur because one (or more than one) driver has a temporary failure and nowadays modern computers are able to simulate rather complex systems (including human behaviors and errors).

This paper, thus, introduces a general simulation framework that can allow the simulation of crashes and the evaluation of consequences on existing microsimulation packages. A specific family of simple and reproducible conflict indicators is proposed and applied to many case studies. In this approach driver failures are simulated by assuming that a driver stops reacting to an external stimulus and keeps driving at the current speed for a given time. The trajectory of the distracted driver vehicle is thus evaluated and projected, for the given time steps, for the established distraction time, over the actual trajectories of other vehicles. Every occurring crash is then evaluated in terms of energy involved in the crash, or with any other severity index (which can be easily calculated since the accident dynamics can be accurately simulated). The simulation of a driver error allows not only the typology of crashes to be included, normally accounted for with surrogate safety measures, but also many other type of typical crashes that it is impossible to simulate with microsimulation and traditional methodologies being caused by vehicles who are driving on non-conflicting trajectories such as drivers speeding at a red light, drivers taking the wrong lane or side of the street or just driving off the road in isolated accidents against external obstacles or traffic barriers. The total crash energy of all crashes is proposed as an indicator of risk and adopted in the case studies. Moreover, the concepts introduced in this paper allow scientists to define other relevant variables that can be used as surrogate safety indicators that consider driving errors. Preliminary results on different case studies have shown a great accordance of safety evaluations with statistical data and empirical expectations and also with other traditional safety indicators that are commonly used in microsimulation.

*Keywords: Traffic simulation; traffic safety; traffic theory.*

*"You progress not through improving what has been done, but reaching toward what has yet to be done."*
**Kahlil Gibran**

---

[*] Corresponding author. Tel.: +39-0984-49-6780; fax: +39-0984-49-6787.
 *E-mail address*: vittorio.astarita@unical.it

# 1. Introduction

*1.1. State of the art in safety evaluation: from traffic conflicts to more elaborated surrogate safety measures and the need for a methodological breakthrough*

The classical methodology to estimate road safety levels is based on applying inferential statistics to crashes databases (Hauer 1986, Jovanis and Chang 1986, Miaou and Lum 1993, Miaou 1994, Shankar et al. 1995, Hauer 1997, Abdel-Aty and Radwan 2000, Yan et al. 2005). With this methodology it is possible to link the causes (infrastructural layouts) to effects (crashes) and apply countermeasures to increase road safety.

In traffic safety (luckily) crashes are a very rare event and sometime new infrastructures black spots could go undetected for years until some very unfortunate event happens. Moreover, a good knowledge of the dynamics of the events that could potentially lead to the actual realization of a crash may provide a more useful base to prevent crashes from happening in the first place, supporting the implementation of appropriate preventive countermeasures.

Surrogate safety performance indicators that provide a causal or mechanistic basis for explaining complex time-dependent vehicle interactions that can compromise safety have been proposed (Hayward 1971; Minderhoud and Bovy 2001; Huguenin *et al.* 2005; Tarko *et al.* 2009).
The first instance of the concept of traffic conflicts was in Perkins and Harris in 1968 that considered the analysis of the use of evasive actions by drivers as an effective method to assess road safety.  The observation of traffic conflicts as temporal and/or spatial proximity started in Sweden with the work of Amundsen and Hyden (1977) and other researchers. Different situations that can arise in traffic streams have been represented by Hyden (1987) in terms of a "safety performance pyramid" (Fig. 1) which characterizes the Swedish traffic conflicts technique.

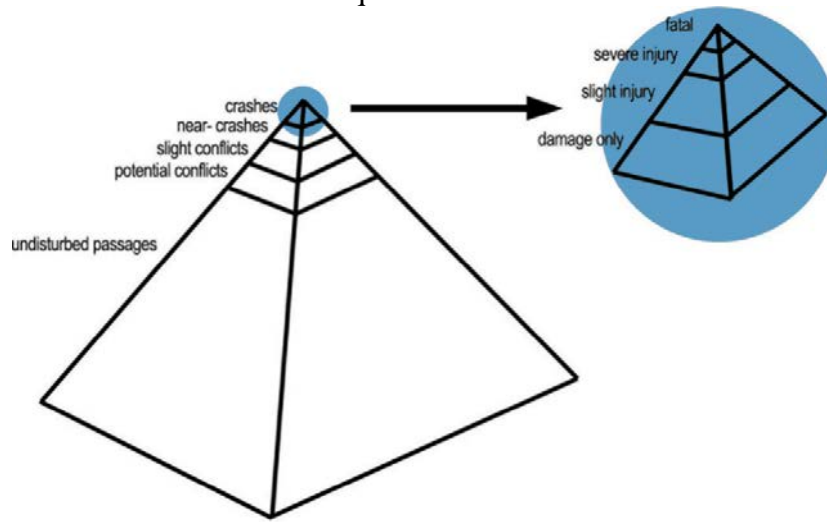

*Figure 1: Hyden "safety pyramid".*

Moreover, traffic conflict analysis originates from a richer data set than statistical analysis based only on crash data since crash occurrences arise from a very small subset of the total set of traffic trajectories data (Fig.2).

Hyden's pyramid represents all possible interactions, ranging from more frequent undisturbed events at the base of the pyramid to less frequent higher risk events nearer the peak (i.e. traffic conflicts and crashes). It is logical to assume that a comprehensive assessment of safety at a given location must reflect the full spectrum of these vehicle interactions since in some "unlucky" cases crashes can generate near the base of the pyramid where conditions are "potentially" safer. Also conventional crash prediction models focus on reported crashes (the very top part of the pyramid), and do not consider unsafe interactions that have not "yet" resulted in reportable crashes.

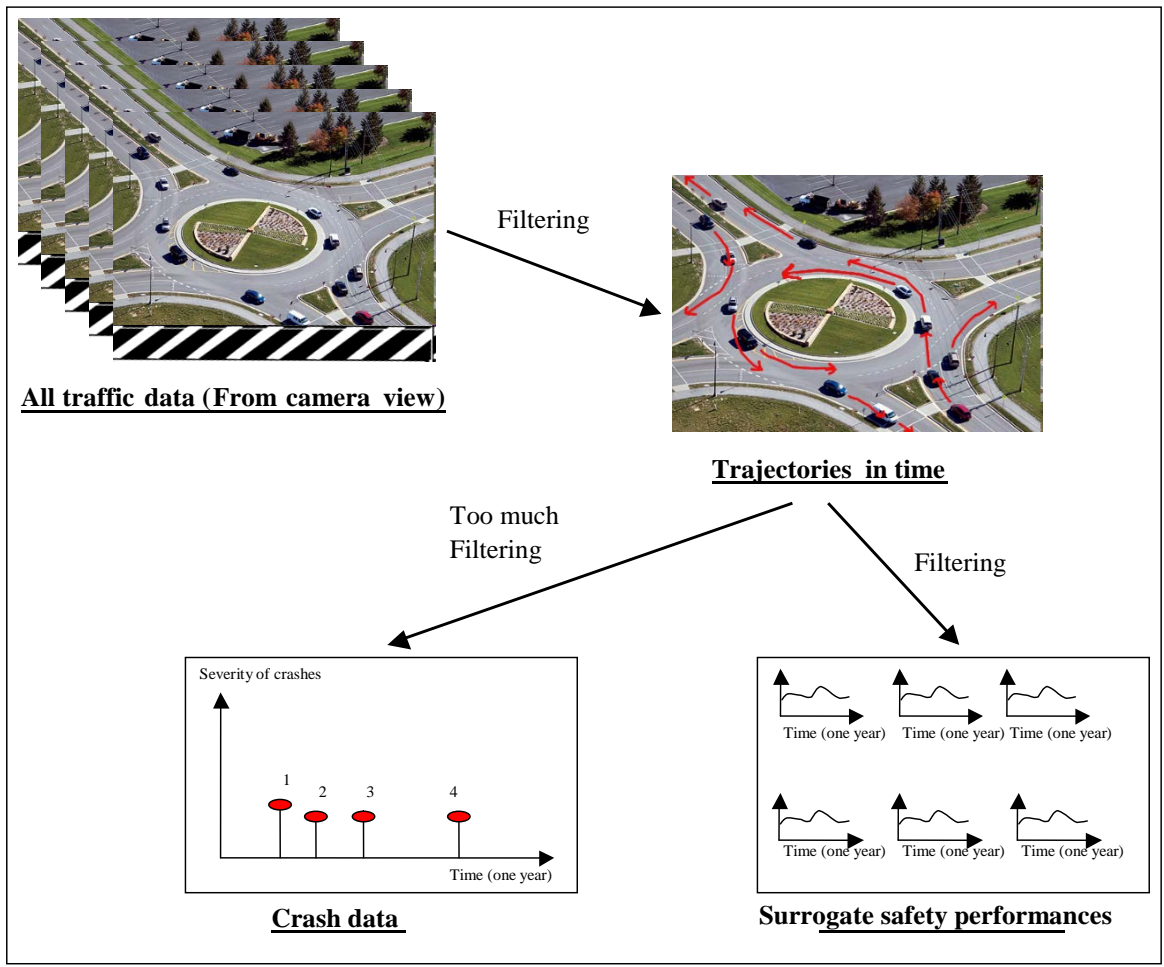

*Figure 2: Both kinematic crash data and safety performances can be obtained from vehicle trajectories: vehicle trajectories data, at a specific site for a given period of time, include also all the trajectories of vehicles involved in a crash (Astarita et al. 2011).*

Unfortunately it is not very straightforward how to make an estimate of the probability of crash events given the frequency and severity of conflicts.

The Swedish traffic conflict observer's manual (Laureshyn and Varhelyi, 2018) says: "If the form of the relation between the severity and frequency of the events is known, it is theoretically possible to calculate the frequency of the very severe but infrequent events (accidents) based on known frequency of the less severe, but more easily observable events (conflicts)". They though acknowledge the fact that the injury risk cannot be easily estimated: "However, it is not obvious how the injury risk in situations where the collision was actually avoided can be estimated."

Many recent works have been directed to establish a relationship between the two different parts of the Hyden "safety pyramid": the white part (where only conflicts occur) and the dark part where crashes occur. Practical problems, posed by the availability of crash data and the methodological challenges caused by the extremely random nature of accidents, have led to the development of simpler approaches to improve road safety assessment, such as the use of microscopic traffic simulation to evaluate traffic conflicts in planned traffic networks.

Traffic computer simulation was introduced in the 50's by Gerlough (1955) and Webster (1958) in 1958 validated his traffic signal delay formula with computer microsimulation. It has been used since then in research to assess the performance of traffic networks and also the introduction of new technologies such as connected vehicles (Astarita *et al.*, 2017).

Current microsimulation models, however sophisticated, do not simulate crashes, the higher dark part of the Hyden "safety pyramid" in Fig. 1 can only be inferred from simulated traffic conflicts (the white part).

Microscopic simulation, applied to establish road traffic safety levels, with the use of traffic conflicts and/or other surrogate safety measures that can be calculated on vehicle trajectories, has been a growing topic of research initially investigated by Darzentas *et al.* (1980) with many following contributions (Cunto and Saccomanno, 2007; Cunto and Saccomanno, 2008; Saccomanno *et al*, 2008; Yang *et al.*, 2010; Cheol and Taejin, 2010; Wang and Stamatiadis,2014). According to the Federal Highway Administration (Gettman et al., 2008), when properly simulated surrogate safety performance measures show the potential to provide a useful platform from which to identify high risk situations in the traffic stream and guide cost-effective intervention strategies. The concept is that the actual risk of crashes can be investigated using traffic simulation sampling of safety performance indicators instead of real traffic data.

Real data of crashes and measured and simulated conflicts have been compared obtaining acceptable results in many papers (Dijkstra et al.,2010; Caliendo and Guida,2012; El-Basyouny et al. 2013; Huang et al.2013; Zhou et al. 2013; Ambros et al. 2014; Shahdah et al.,2014,2015). Microsimulation and surrogate safety performances have also been used to assess the impact on safety of the introduction of connected and autonomous vehicles (Zha et al., 2014; Morando et al. 2018, Astarita et al. 2018 )

The analysis of conflicts and the use of traffic simulation for the evaluation of road traffic safety levels has accelerated research efforts into establishing when two vehicles trajectories constitute a "traffic conflict". Many "surrogate safety measures" have been proposed such as: time to collision (TTC), post encroachment time (PET), initial deceleration rate (DR), maximum of the speeds of the two vehicles involved in the conflict event (MaxS), maximum relative speed of the two vehicles involved in the conflict event (DeltaS), deceleration rate to avoid the crash (DRAC), and proportion of stopping distance (PSD).

Almost all of the proposed "surrogate safety measures" works on trajectories of vehicles (that will not crash into each other) to establish a number of potential conflicts arising from temporal or spatial proximity under the assumption that the closer vehicles are to each other, the nearer they are to a collision (Mahmud *et al.*, 2017). The proximity is established (mainly with thresholds) on the basis of quantitative measurements that are obtained from vehicle trajectories.

The interested reader on this subject can refer to three thoroughly redacted (and relatively recent) papers that give a state of the art respectively on traffic conflicts, road safety simulation modeling and on the use of proximity surrogate safety measures (Mahmud et al., 2017; Young et al. 2014; Zheng et al. 2014)

All the three above-cited "state of the art" papers evidence advantages and also some limitations of using microsimulation and surrogate safety measures.

Zheng *et al.*(2014) raise two conceptual issues: process model validity and severity inconsistency. Process model validity originates from the fact that some researchers believe that traffic conflicts and crashes are two mutually exclusive events while others believe that traffic conflicts could lead to traffic crashes where no evasive driving countermeasure is taken. Severity inconsistency is an issue that originates from the fact that the severity of traffic conflicts is identified either by the intensity of evasive actions to avoid a crash or by the proximity in time (or space) between vehicles while severities of crash events cannot be measured in the same way. In fact, severity of crash events is usually expressed in terms of severity of consequences.

Zheng *et al.*(2014) say about crashes that: "proximity measure can only reflect the risk of collision but not the severity of crashes, which is typically differentiated based on consequences of the crash not proximity."

Young *et al.*(2014) conclude their work by recognizing that simulation will become a useful tool for road traffic analysis, yet they write that there are a number of areas where further research is needed:

- "The crash as the measure of performance": They recognize that models needs "theoretical and numerical improvements" to specify the factors that originate them. The relationship between traffic conflicts and the real crash dynamic has not been adequately investigated.
-"The theory behind driver behavior in crashes": Driver behaviors that results in a crash have not been taken into account holistically in simulated environments.
-"A more detailed representation of the vehicle and conflict situations": Surrogate safety measures are based on trajectories of a point that represents the barycenter of a vehicle, real vehicle dimensions and physical interactions at crashes are not commonly considered.
-"A generalization of the models to look at more crash and vehicle types": Different type of vehicles such as heavy vehicles have not been specifically considered.
Laureshyn *et al.* (2017) clarify that commonly used conflict indicators are able to express "proximity" to a crash and neglects the potential severity of the consequences of a crash. The consequences need to be taken into account in some way (Laureshyn *et al.*, 2010).

Moreover, the mechanism that generates a crash has not been clarified. In traditional approaches there is no clear explanation of how a near-crash situation can become a real crash. It is often assumed that "proximity" of vehicles that are in a near-crash event means also a higher probability of crash.
What are the causes of a real crash? What needs to happen to make a near-crash event turn into a crash? What are the causes of random deviations from a safe trajectory that make vehicles crash into other nearby vehicles? What is the mechanism that creates a jump from one region to another region of the Hyden "safety pyramid" (see figure 3)?

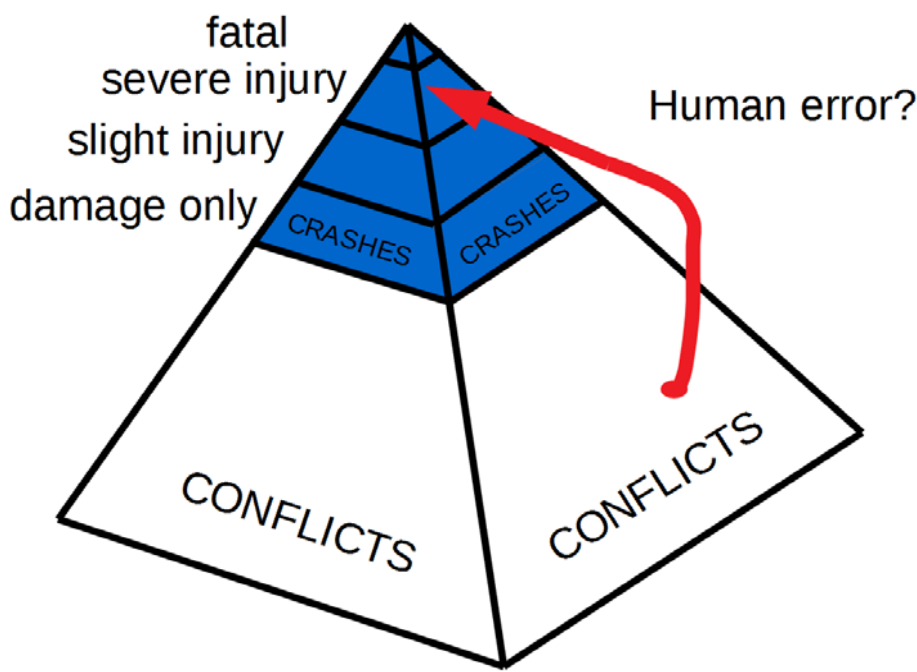

*Figure 3: Unexplained discontinuity between near-crashes and real crashes.*

Davis et al. (2011) outline a general causal model of traffic conflicts and crashes based on a probabilistic approach, we believe that it is important to take into account the human factor and this has been seldom taken into account in preceding works.
Bevrani and Chung (2012) recognize the importance of human error since according to Stutts et al. (2003) 30 % of police reported crashes in the USA are due to human distractions. Stutts et al. (2003) found that some types of distracting behavior that people undertake while driving include talking on a cell phone, shaving, putting on eye makeup, eating or drinking, changing a radio station or compact disc and looking at the back seat to talk to an infant in a car seat.
We conclude this brief state of the art with the reflection that, possibly, the two main factors that have not been thoroughly investigated in traffic conflict simulation are the potential consequences of crashes and the human factor.

This paper is a first attempt to evaluate traffic safety with microsimulation introducing possible driver errors. The paper shows that introducing driver error is a feasible task by using common microsimulation packages and some of the above-indicated issues could possibly be addressed given a breakthrough in research that could properly introduce driver error in the simulation and shift the focus from traffic conflict simulation directly to traffic crash simulation (raising awareness of potential consequences).

*1.2. Specific problems in the simulation of conflict indicators*

Microsimulation of conflict indicators (instead of measuring them on the field) is more controversial since microsimulation, instead of direct observation, could exacerbate some of the issues indicated above. Specific problems are listed in the following:

**- Process model validity.** Commonly used safety indicators in microsimulation are numerically based on variables which, in most cases, have nothing to do with the real dynamics of real accidents. As an example, at the present state of the art, there is no way to simulate or take into account accidents that are originated by specific careless driver behaviours such as vehicles speeding at the red light of a traffic signal or in the general case of vehicles which are travelling on non-conflicting trajectories. Many real crashes are generated by obvious driver mistakes and current microsimulation models have no way to simulate this.

**-Traffic simulation packages do not explicitly simulate safety indicators.** SSAM software was developed to be applied as an add-on on to main traffic simulation packages (Pu *et al.*, 2008). Some of the problems with the use of SSAM software that are evidenced in Souleyrette and Hochstein (2012) are the lack of severity classification of conflicts and the lack of ability to map risk and selected conflicts. Souleyrette and Hochstein (2012) suggest many possible modification to SSAM software.

**-Traffic safety indicators have been developed to be used with real trajectories and traffic microsimulation is specifically designed to avoid crashes.** Surrogate safety measures have been developed to be used with real trajectories. The use of microsimulation to evaluate surrogate safety measures is questionable since traffic microscopic models tend to move vehicles around the network avoiding any interaction that could lead to a crash. Crashes detected by SSAM software are, in fact, caused by graphical inaccuracies of microscopic simulation packages. Gettman *et al.*, (2008): "All of the simulation systems exhibit modelling inaccuracies that lead SSAM to identify conflict events with TTC = 0 ("crashes")".

**-Traffic microsimulators need calibration.** The use of traffic microsimulation to obtain trajectories on which surrogate safety measures are calculated is subject to a correct determination of input parameters. Microsimulation calibration is a delicate matter and it is not guaranteed that even after performing a perfect rigorous calibration of a microsimulation model (on traffic dynamic parameters) the trajectories will produce estimates of safety performance that can be matched with real world observations. For this reason calibration based on safety performance measures has been proposed (Cunto and Saccomanno, 2008) though, there is no guarantee that it will lead to a calibration which is consistent with traffic dynamics.

**-Traffic Safety Indicators based on conflicts (TSI) are numerous and there is no consensus on which one to use.** Some of them are specifically designed for some specific manoeuvres, moreover, the commonly used TSI do not consider crash severities. It is not easy to choose among safety indicators since, as stated above, the complexity involved in the relationship between potential conflicts and real crashes is not straightforward. In Tarko et al. 2009 it is said: "Consider excessive braking maneuvers. One might presume that observing these - may serve as a reliable predictor of crashes. However, some drivers who apply brakes aggressively are avoiding crashes, while others who fail to apply brakes are involved in

crashes. These two simple examples illustrate in a simple and perhaps naïve way the potential complexities involved in selecting reliable safety surrogates."

**Most Traffic Safety Indicators (TSI) do not consider real crash dynamics and potential consequences.**
Most surrogate safety measures commonly used are able to establish conflicts and potential crash yet they are not able to make a differentiation between different severity of crashes. Crash severity is an extremely important measure (Ivan *et al.*, 2018) yet the commonly used conflict safety indicators consider all possible collisions incidents as having the same severity and this does not allow accounting for the fact that accidents occurring at a higher relative speed between vehicles have more severe consequences. Only recent research efforts have been directed to addressing this issue (Laureshyn *et al.*,2017).

**Friction and shear forces effects on safety are not considered in traffic flow theory or in any conflict indicator.**
Conflicts between vehicles travelling on trajectories that are not intersecting are never considered in common conflict indicators. Drivers drive all the time very close to roadside obstacles or barriers and also very close and often to vehicles coming from the opposite direction. In all these very common situations a very temporary and small deviation from the right trajectory would cause a crash. These types of conflict are not at all considered in common conflict indicators and this is a serious inadequacy that has not been explored in science. This point is addressed in the following presented microsimulation case study (second case).

The proposed methodology can take into account isolated vehicle crashes with lateral barriers or objects (these events cannot be taken into account with commonly used TSI) and also give a solution to some of the above-listed problems.
Especially crash dynamics can be reproduced accurately with microsimulation and this potential has not been explored. Inelastic and elastic collision hypothesis could be applied to obtain a better estimate of the consequences of a crash. For this reason, in the next section, simple crash dynamics (dating back to Newton laws) in vector form is presented. It also serves as a base for the proposed methodology.

*1.3. Crash severity in simulated surrogate safety measures and inelastic collision hypothesis*

It is possible to correct or extend commonly used safety indicators to take into account the amount of energy involved in near-crashes events. In fact, the severity problem in simulated surrogate safety measures was addressed in Shelby (2011,with the introduction of Delta-V as a measure of traffic conflict severity), Sobhani et al. (2011,2013, with the use of kinetic energy) and in Laureshyn et al. (2017, with the introduction of the use of extended Delta-V). The values that are involved in estimating crash severity are the speeds and the masses of colliding vehicles.
The dynamic of a crash involving two vehicles can be studied with a fundamental law of physics. The laws of physics that are applied are conservation of energy and momentum.
For a monodimensional impact, the relative velocity of the bodies after the impact is proportional to that preceding the impact through a restitution coefficient linked to the elasticity of the two bodies (Newton):

$$v_{1a} - v_{2a} = -\varepsilon(v_{1b} - v_{2b}) \qquad (1)$$

with:

$$0 \leq \varepsilon \leq 1 \qquad (2)$$

where $v_{1a}$ and $v_{2a}$ are the speeds of objects 1 and 2 before the collision, if $\varepsilon = 0$ the impact is called totally inelastic; if $\varepsilon = 1$ the impact is called elastic.
The difference between a perfectly elastic or perfectly inelastic collision is in terms of how much kinetic energy is adsorbed in the collision by the two body colliding. A perfect conservation of energy (almost impossible) corresponds to an elastic collision while the maximum adsorption of energy corresponds to a collision where the maximum possible energy is absorbed and where both objects will stick together and move at the same speed after collision. Real car crash collisions as noted in Shelby (2011) are somewhere between elastic collisions and inelastic collisions. A purely elastic collision would be when two cars after touching each other bounce back without deformation (and without kinetic energy loss) while a perfectly inelastic collision would be when there would be no bouncing and the car would form, after the collision, a single body where a part of kinetic energy has been absorbed into the crash and the rest would be still present in the motion after collision. Shelby (2011) citing Nordhoff (2005) affirms that car collisions tend to have a coefficient ε of restitution of about 0.4 at low-speed for bumper-to-bumper collisions, reducing to around 0.1 for higher speed collisions where the vehicle bodies would deform substantially. It is very convenient for approximate calculations such as those used in surrogate safety measures to consider car crashes as inelastic. Shelby (2011) notes how the original CRASH program commissioned by the National Highway Traffic Safety Administration NHTSA to estimate initial impact speeds assumed an inelastic collision and that this his was later identified as a reason for underestimating speed values by 10% to 30% (McHenry and McHenry,1997). A subsequent version of the CRASH software was updated taking into account elastic behavior and initial impact speeds were estimated as close as about 1% of their true value.

We will shortly synthesize two vehicles crash dynamic laws in vector notation under the assumption of inelastic collision since, in the remaining part of the paper, we will develop crash severity indexes that are based on this assumption. It must be noted that the proposed methodology in future developments could take into account partially elastic collisions as we believe that elastic conditions can play an important role in crash dynamic situations where multiple impacts are involved and where the first low energy impact could cause more severe secondary collisions (for example: a car which hits laterally another car running in the same direction could cause the second car to hit a lateral obstacle or invade opposite traffic lanes or in crashes that involve more than two vehicles).

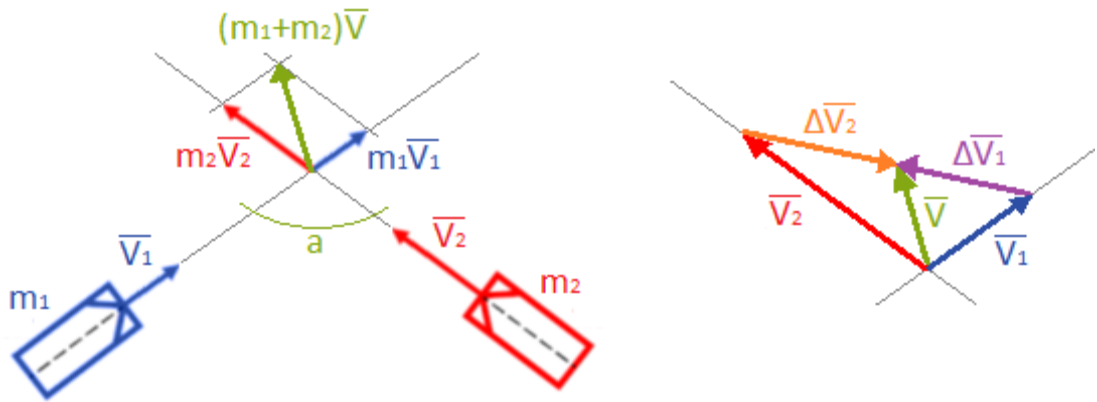

*Figure 4[1]: Calculation of the resulting speeds: $\overline{V}$, $\overline{\Delta v_1}$ and $\overline{\Delta v_2}$ based on momentum conservation principle (both elastic and inelastic collisions).*

The law of conservation of momentum states that the momentum $\overline{P}$ (a two dimensional vector in our case) expressed as the product of mass and velocity keeps constant after a collision (both elastic and inelastic), since the momentum $\overline{P}$ before the collisions is:

$$\overline{P} = m_1 \cdot \overline{v_1} + m_2 \cdot \overline{v_2} \qquad (3)$$

where $m_1$ and $m_2$ represent the masses of the two vehicles and $\overline{v_1}$ and $\overline{v_2}$ are the velocities (vectors) before the impact, the velocity $\overline{V}$ of the entire system after the collision can then be calculated as :

---

[1] This image corrects some minor typographical inaccuracies in the images presented in Shelby (2011), Jurewicz *et al.*(2016) and in Laureshyn *et al.* (2017).

$$\overline{V} = \frac{m_1 \cdot \overline{v_1} + m_2 \cdot \overline{v_2}}{m_1 + m_2} \qquad (4)$$

Delta-V is a notation often used to denote a vehicle change of velocity experienced during a crash. Delta-V has been statistically connected with the level of injuries passengers can suffer (Gabauer and Gabler 2006; Tolouei *et al.* 2013) and in our two dimensional example is given by:

$$\overline{\Delta v_1} = \overline{V} - \overline{v_1} \text{ and } \overline{\Delta v_2} = \overline{V} - \overline{v_2} \qquad (5)$$

for the two vehicles involved in a crash.

By definition we have that:

$$\overline{\Delta V_{12}} = \overline{v_2} - \overline{v_1} = \overline{V} - \overline{\Delta v_2} - \overline{V} + \overline{\Delta v_1} = \overline{\Delta v_1} - \overline{\Delta v_2} \qquad (6)$$

Moreover as shown in figure 4 $\overline{\Delta v_1}$ and $\overline{\Delta v_2}$ are always parallel and opposite vectors, $\overline{V}$ being a linear combination of $\overline{v_1}$ and $\overline{v_2}$.

The Kinetic energy of the two vehicles before the impact is:

$$K_a = \frac{1}{2} m_1 v_1^2 + \frac{1}{2} m_2 v_2^2 \qquad (7)$$

where $v_1$ and $v_2$ are the modules of the two velocity vectors. The Kinetic energy of the two vehicles after the impact under the hypothesis of inelastic collision is (vehicles stick together and keep same speed):

$$K_b = \frac{1}{2}(m_1 + m_2)V^2 \qquad (8)$$

where $V$ is the module of the common speed vector after the crash. By developing the difference in Kinetic energy before the crash and after the crash, it is possible to calculate the energy that is absorbed by the two vehicles in the impact that can be expressed as:

$$\Delta K = K_a - K_b = \frac{1}{2} m_1 v_1^2 + \frac{1}{2} m_2 v_2^2 - \frac{1}{2}(m_1 + m_2)V^2 = \frac{1}{2} m_r (\Delta V_{12})^2 \qquad (9)$$

where $\Delta V_{12}$ is the module of the difference vector between the two impact velocities $\overline{v_1} - \overline{v_2}$, $K_a$ and $K_b$ are the kinetic energies before and after the crash and :

$$m_r = \frac{m_1 m_2}{m_1 + m_2} \qquad (10)$$

is called the reduced mass of the system.
In all calculations that are presented in the following, one of the simpler and synthetic safety indicators that is used is the total energy of a scenario which is evaluated as the sum of ΔK for all simulated crashes.
Since the proposed methodology is applied as an add on over traffic simulation a more detailed analysis can also be performed in which the energy absorbed by every vehicle and the exact vehicle type is considered.

In that case, it would be useful to use the formula proposed by Joksch (1993), as already done in Evans (1994), that directly establishes a probability of death or injury as a function of Delta-V:

$$P = \left(\Delta V / \alpha\right)^k \qquad (11)$$

For this more detailed calculation involving the probability of death or injury the quantity of energy absorbed by each vehicle can be calculated as:

$$\Delta K_1 = \frac{1}{2} m_1 |\overline{\Delta v_1}|^2 \ , \Delta K_2 = \frac{1}{2} m_2 |\overline{\Delta v_2}|^2 \tag{12}$$

It must be noted that it is easy to demonstrate that the total energy absorbed by the vehicles $\Delta K$ can be assigned in equal parts to the two vehicles when they have equal masses:

$$\Delta K_1 = \Delta K_2 = \frac{\Delta K}{2} \tag{13}$$

In the general case, we have:

$$\frac{|\overline{\Delta v_1}|}{|\overline{\Delta v_2}|} = \frac{m_2}{m_1} \tag{14}$$

This latter equation explains why when the vehicles have different masses the lighter vehicle will suffer greater damage and its occupants face a higher probability of death or injuries according to (11).

*1.4. Contribution of the paper as a breakthrough in traffic simulation: introducing driver error and "potential crashes" instead of looking for "potential conflicts" in existing simulations*

In this paper, a completely new approach to microscopic evaluation of surrogate safety measures is proposed. The proposed safety indicator is based on the explicit simulation of possible driver errors. The idea originates from the consideration that, in the real world, most accidents involving human drivers occur because one (or more than one) driver has a temporarily distraction.

A driver for example can be distracted by being occupied in some other tasks (for example a mobile call) or because he has a momentary breakdown in attention and/or awareness of external changes due to a psychological or physical failure. There are many driver distractions that could cause human failure as revealed by Stutts et al. (2003). In Ueyama (1997) on the base of police records it is shown that in many right-angle collisions at an intersection in Tokyo drivers did not attempt to avoid the crash.

Bevrani and Chung (2012) propose a framework for car-following microsimulation models (Figure 5) to include human errors. They identify three stages in which a human error can take place: perception and recognition, decision making, dynamic formulas.

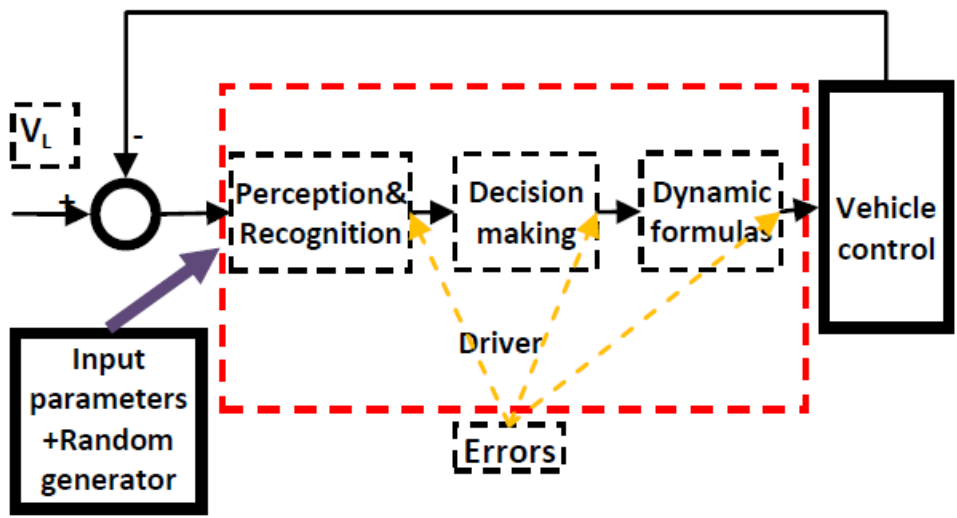

*Figure 5: Bevrani and Chung (2012) framework to introduce human errors in microsimulation models.*

The simulation and study of crashes and consequences has been carried out in many works (Mak *et al.* 1998; Archer and Kosonen,2000;Erbsmehl,2009) although it has never been carried out in a complete simulated traffic scenario such as one that is produced by microsimulation.

In this paper, a first methodology to shift from "potential conflicts" simulation to "potential crashes" simulation is proposed, by introducing human failure in terms of potential driver errors, and described in the next section.

A complete microsimulation application has been developed that is able to explicitly simulate potential driver errors. The microsimulation model that has been developed in this work has been implemented as an add-on to the existing Tritone microsimulation model (Astarita *et al.*, 2011; Astarita *et al.* 2012) and can be applied also to other existing microsimulation models such as Vissim (PTV,2005) or Aimsun (Barcelo *et al.*, 1994).

In the following case studies, driver failures are simulated by assuming that a driver stops reacting to external stimulus and keeps driving at the current speed for a given time (we have called this driver a "Zombie" driver and the vehicle which is driven by a "Zombie" driver a "bullet" vehicle). No decision making and dynamic formulas that would account for an evasive maneuver (such as an emergency braking) are taken into consideration in the case studies presented in this paper. Moreover, in our simple approach also other drivers are not able to perceive and react to the conflict that a "Zombie" driver could be generating.

## 2. The proposed methodology

One of the criticisms that is often raised to the application of microsimulation for the simulation of crashes is the fact that crashes are very random and rare events and thus it is not possible to reproduce them. In our methodology, we assume that crashes are originated by human errors. Driver errors happen more often than crashes and not all driver errors have as an outcome a crash.
The first problem to deal with is the rate of distraction and how a specific distraction turns into a specific driving error. Let us assume that a distraction turns out into a specific erroneous driving pattern. The rate of distraction (or erroneous driving patterns) $\lambda$ can be assumed as a variable which measures the number of distractions that a driver can have while driving for a given amount of time. The rate $\lambda$ is not enough to characterize the distraction phenomenon. A distribution $\psi$ of the typology of erroneous driving pattern (or severity of the distraction) must also be assumed. In theory, knowing the rate $\lambda$ and the distribution of $\psi$ a Monte Carlo method could be applied to generate randomly the instant of distraction and the consequent erroneous driving pattern.

An underlying assumption for this rate-of-distraction-based model could be that crash frequency is connected to the distraction frequency, which can be assumed to occur randomly and uniformly along any travelled distance. In other words, a driver has an equal probability of making an error at any point of his trip. However, crash data have shown that crash frequencies are higher at certain locations and in certain traffic flow conditions. How to deal with these complications?

A very similar methodology, to what is proposed in the following, has been applied into the Roadside Safety Analysis Program (RSAP) developed in the National Cooperative Highway Research Program (NCHRP) and presented in report 492 that is the Engineer's manual for RSAP (Mak and Sicking 2003). The NCHRP report 492 is more than a software manual and describes the basis of a research project to develop an improved cost effective analysis procedure for assessing roadside safety improvements. The project objective has been accomplished with the development of a software that can help roadside safety engineers to evaluate impacts of roadside safety improvements. The RSAP software consists of four basic modules: the Encroachment Module, the Crash Prediction Module, the Severity Prediction Module, and the Benefit/Cost Analysis Module. Using the Monte Carlo technique, RSAP generates the following conditions: encroachment location, encroachment speed and angle, vehicle type and vehicle orientation on impact. RSAP assumes a straight encroachment vehicle path justified by "the lack of quantitative information on vehicle and driver behavior after encroaching into the roadside" and also due to the difficulty of simulating a curvilinear vehicle path. In other words, RSAP generates randomly (in a random location) a vehicle and its roadside encroachment trajectory (starting from a random location on the road) and assesses the consequent damage of a "potential" crash by evaluating the consequences of simulated impacts.

The proposed methodology differs from the RSAP software approach since RSAP considers only isolated driver crashes against roadside obstacles and barriers without considering complete traffic scenarios and interactions among vehicles. Moreover, RSAP is not built upon traffic microsimulation software.

The proposed methodology is instead built upon classical microsimulation packages. Knowing the vehicle involved in a distraction and the consequent severity of the distraction that leads to an erroneous driving pattern it is possible to create an "add-on" simulation inside a classic simulation scenario. The "bullet" vehicle kinematics can then be studied along with that of other surrounding vehicles. A similar procedure has been applied in the paper of Wang et al. (2017) where three anomalous driving patterns are considered: vehicles which are moving at an excessive low or high speed and vehicles that operate an emergency braking maneuver that makes them abruptly stop. In that work, an "add-on" simulation is performed only on the chosen anomalous vehicle and on neighboring vehicles.

Our methodology is different from the method proposed in Wang *et al.* (2017) since we introduce explicitly driver distractions as a possible anomalous behavior that turns into an anomalous driving pattern. Moreover, we propose a detailed analysis of crash dynamics that considers also isolated crashes with roadside obstacles and barriers. Apart from that, the general approach is very similar. And the third case of sudden stops analyzed in the paper of Wang *et al.* (2017) can be seen as a subcase of road failure in our general methodology.

In our methodology, we assume that the outcome of a given specific erroneous driving pattern can be easily studied by applying vehicle kinematic in a microsimulation program and that this "add-on" simulation can be performed with a post-processing procedure that is performed on the trajectories that are the output of a classical microsimulation.

This "add-on" simulation in general could be as simple as projecting the "bullet" vehicle along a straight trajectory line as in RSAP or as complicated as a full microsimulation model. The general framework, that we believe is applicable to every kind of traffic anomaly, is such as depicted in Figure 6. In Fig. 6 we introduce also the term "road anomaly" since a vehicle which operates an abrupt stop most of the time is forced to do so because of some kind of road anomaly (an obstacle on the roadway).

Driving behaviors (or road) anomalies can be simulated by using a random Monte Carlo approach or a deterministic approach. In the first case, parameters affecting perception, decision making and vehicle dynamics (in the case of failing vehicle) are randomly generated. In the deterministic approach instead sets of parameters are fixed and the calculation must be repeated for every set of parameters. It must be noted that road failures such as a dog crossing the road or exceptionally slippery road pavement (due for example to an oil spill) can also be simulated with this approach. The start of the procedure in Fig.6 is the choice of a vehicle (driver) who performs an anomalous driving pattern. This pattern is simulated together with the potential evasive movements of other surroundings vehicles. The output of this simulation, if a crash occurs, is considered in terms of severity. We propose to use general simple energy-based indicators for severity of a crash, although every detailed analysis of the crash dynamics can be performed.

In this paper, we apply a simple methodology that can be considered a first simple implementation of the general framework of Fig. 6 based on a deterministic technique. A Monte Carlo or mixed deterministic-Monte Carlo approach could also be used, though, since this is the first implementation of this methodology, we have kept this first implementation simple and reproducible by applying a deterministic approach instead of a stochastic approach.

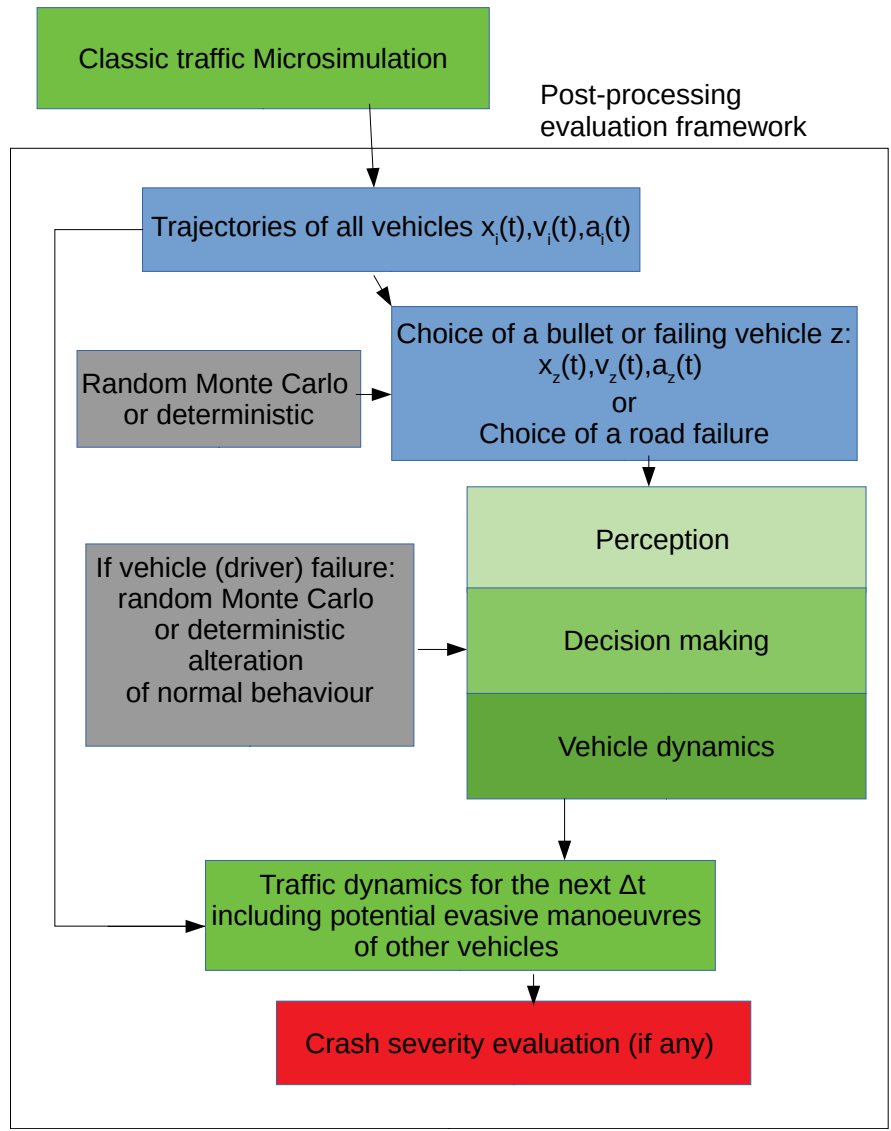

*Figure 6: General proposed framework for the simulation of driver, vehicle and/or road failures.*

In general terms, the proposed first implementation of our methodology develops in a very similar way to RSAP, applying the same procedure of projecting the trajectory of a vehicle in a straight line. Yet we propose to apply this over the simulated trajectories of other vehicles in a microsimulation environment scenario. In other words, the trajectory of the distracted driver vehicle is evaluated and projected, for the given time steps, for the established distraction time, over the actual trajectories of other vehicles. Every occurring crash is then evaluated in terms of energy involved in the crash. The simulation of a driver distraction is performed every time step on every vehicle. With this methodology a great number of "potential" crashes are then generated. The total crash energy (sum of all energy involved in all potential crashes) or other relevant variables can then be used as safety indicators.

In detail, the framework we applied in our case studies is depicted in Figure 7. There are just three main parameters of our first implementations: the simulation time step, the distraction time and the angle of the possible "Zombie" vehicle trajectory. Each single vehicle that is moved on the network is considered as candidate for a driver distraction each time step of the microsimulation. The time step, so, is the first parameter of the proposed methodology and in this first implementation was set equal to one second. In the following case studies, microsimulation of traffic scenarios is used as a basis for calculations, and in each microsimulation, each vehicle on the road is considered every second for a "potential crash" production. A "potential crash" is generated by moving the vehicle (starting at time $t_0$ ) for a given distraction time $\Delta T$ (until the time $t_0 + \Delta T$) at a constant speed equal to the simulated speed of the vehicle at time $t_0$. This second parameter, $\Delta T$ in this paper, has been set as an integer value from one to seven seconds. In this preliminary work, to make our work reproducible we have used fixed values instead of a random distribution with a Monte Carlo technique.

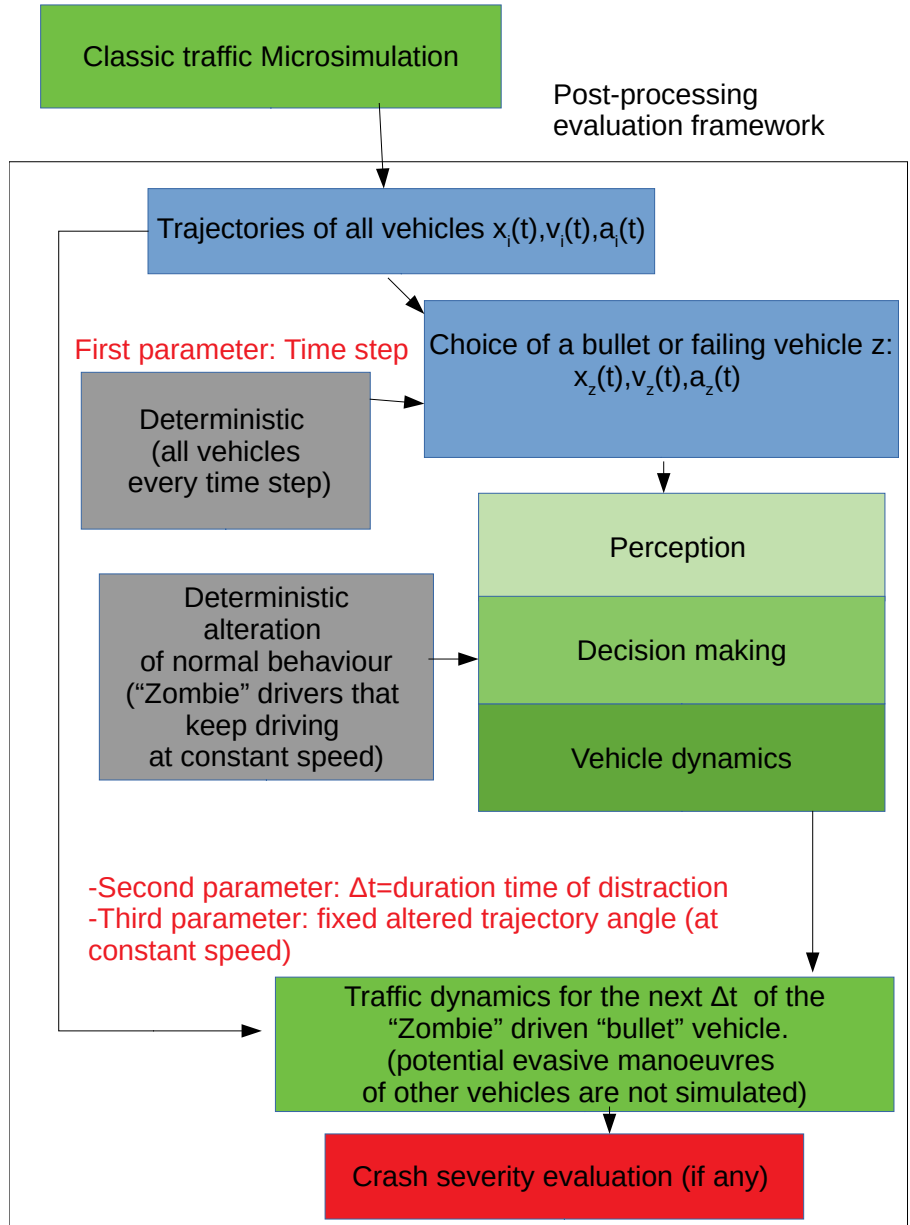

*Figure 7: Proposed specific framework for the simulation of driver errors following the "zombie" driver ("bullet" vehicle) approach. This framework has been used in the following microsimulated case studies and is a subset of the general framework in Figure 6.*

Also the trajectory of the vehicle could be simulated at a random angle generated with a Monte Carlo technique according to a given distribution. In this first implementation instead we used 3 different angles: a straight line, according to the given direction of the vehicle at time $t_0$, and a 15 degrees deviation on the left and on the right.

In other words, as an example, for a parameter ΔT of 5 seconds for each vehicle we calculate 15 vehicle positions as depicted in Figure 8.

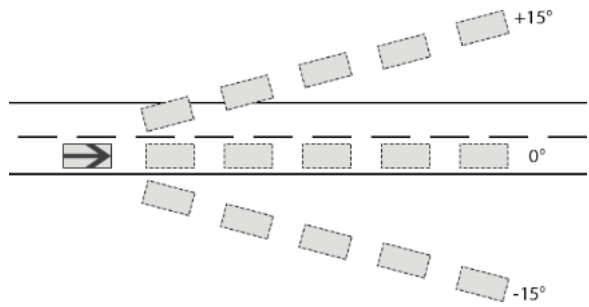

*Figure 8: Possible positions of vehicle trajectories (as they are investigated in the following case studies) with a time distraction of 5 seconds.*

In the following simulation case studies, collisions between vehicles and roadside objects or barriers have not been considered so the crash depicted in Figure 9 is possible (with DT=7 sec., it must be noted that a lower DT would not have caused a collision).

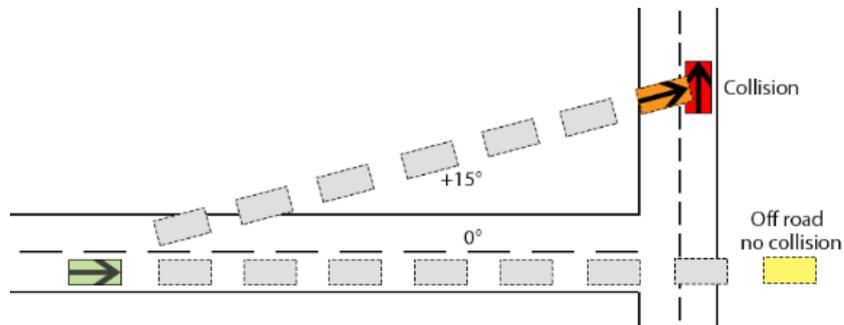

*Figure 9: A rare but possible collision at a 15 degrees angle trajectory after 7 seconds of distractions.*

The detail algorithm of the microsimulation add-on examines all vehicle trajectories every second of the simulation, performing a "distraction" simulation inside the main simulation. Each vehicle, which is moving at the considered time, is advanced at a constant speed, for an imposed period of time on a deviation angle that is imposed for the direction. During this subsimulation it is possible to have two results: the vehicle does not impact with any other vehicle (barrier or side obstacle) or the vehicle is involved in a crash. The collision between two vehicles is detected by evaluating the points of intersection between two vehicles as in Figure 10. The algorithm evaluates whether the sides of a vehicle intersect with the sides of another vehicle (or barrier).

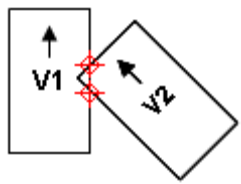

*Figure 10: Collision detection.*

If there is a collision the subsimulation is interrupted and the extent of the collision is evaluated as a function of the energy produced. In the end, the algorithm returns detailed information on individual events and aggregate statistics, which allow a rapid evaluation of the analyzed network. Aggregated information (averages, variances, sums, minimums and maximums) can be calculated on the basis of the number of collisions, the effective duration of the distraction, the overall energy developed during the collision, and the amount of energy for each of the two vehicles involved.

An accurate analysis of road side barriers safety performances is beyond the scope of this paper so only an analytical example is presented in the following section. All the presented simulation-based case studies (similarly as in all preceding papers based on the evaluation or microsimulation of surrogate safety measures) do not consider isolated vehicle crashes or road side crashes. It must be noted though that the proposed methodology allows researchers also to assess easily the real impacts of roadside safety improvements basing analysis on the real prevalent time varying o/d matrix in every traffic scenario that can be reproduced by microsimulation.

## 3. How the three parameters affect results, under the proposed specific framework, in the case of isolated vehicles: two simple examples analytically evaluated

In the following, a special notation will be used to indicate parameters of a surrogate safety indicator such as that proposed in the previous section. The parameters are the distraction time duration, the angle of the deviated trajectory and the percentage of energy considered for a straight deviation. As an example, for a distraction time of 3 seconds, a trajectory deviated on the left and on the right at an angle of 15° and an energy calculated by summing up 80% of the energy due to crashes that are originated by a straight trajectory (10 % for the energy of the left and right deviated trajectories) would be indicated by the indicator: $Z_{3\text{-}15\text{-}0.80}$. This means that $Z_{3\text{-}15\text{-}0.80}$ of a given traffic scenario would be the sum of 80% of all the energy relative to all crashes that are generated by straight trajectories (of 3 seconds) and 10% of the sum of all energy of crashes that are generated by trajectories that deviate left and right at an angle of 15°.

*3.1 Example 1 isolated vehicle and linear road side barrier.*

The first proposed example (see Fig.11) is a single vehicle travelling along a road segment of one km at a constant speed of 90 km/h, the road side barrier is a concrete linear wall.
If we make the case that diver failures are generated every second (first parameter time step = 1 sec.), that the driver distraction time is 5 seconds and we consider equally the energy coming from the three possible angular trajectories above indicated (deviation angle +-15° which means that the energy calculated is : $Z_{5\text{-}15\text{-}1/3}$). Then two crashes are generated by each one of the 40 considered positions of the vehicle. The two crashes at a speed of 90 km/h against the vertical walls on both sides of the road happen as a consequence of the 15 degrees deviation to the left and to the right. For the crash severity evaluation, we can consider the component of kinetic energy in the direction perpendicular to the wall:

$$\Delta K = \frac{1}{2} m (v_x)^2 \qquad (15)$$

where $v_x$ is equal to 90*sin(15°)=23.3 (km/h)=6.47 (m/s) which, if we make the case of a vehicle of 1000 kg, brings a result of $\Delta K$= 20 933 (Joule) for the lateral crash. Considering the 40 positions of the vehicle in one km of road and two crashes for every position this brings a total crash energy of 1 674 682 (Joule) (which must be divided by three according to the above-described indicator). This example clearly shows that the total crash energy would be inversely proportional to the chosen time step.
Changing the trajectory angle by increments would also increase the energy value. The general formula for the total crash energy, in example one, as a function of lateral angle deviation and time step is :

$$\text{Total crash Energy} = Z_{5\text{-}\theta\text{-}1/3} = \frac{L}{v \cdot \delta t} \cdot 2 \cdot \frac{1}{2} m (v \cdot \sin \theta)^2 * 1/3 \qquad (16)$$

where $\delta t$ is the time step (s), $v$ is the vehicle speed (m/s), m the vehicle mass (kg), θ the deviation angle and L the road segment length. An increase in the distraction interval ΔT would not affect results of road side crashes, since in this example only one vehicle is on the road, while instead it would obviously affect total crash energy in all scenarios where different vehicles are simulated in a microsimulation scenario. A decrease in ΔT could affect results, since a small ΔT would generate crashes only against barriers or objects that can be reached by the "bullet" trajectory.

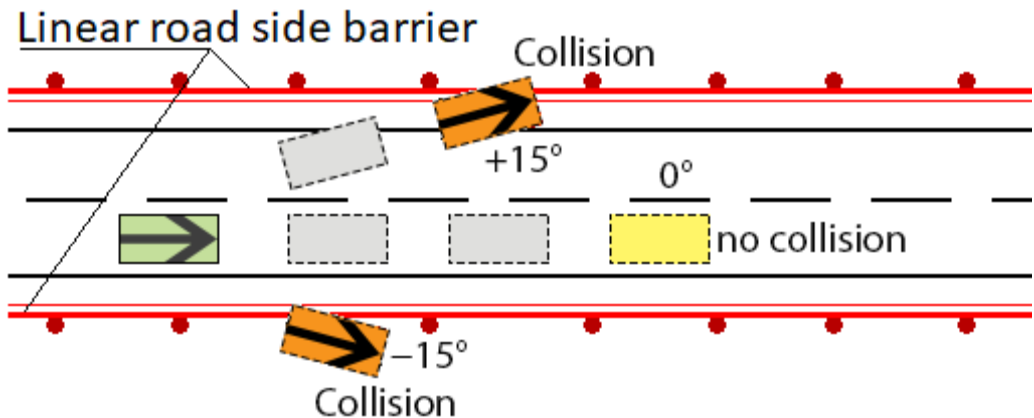

*Figure 11: Example 1.*

*3.3 Example 2 isolated vehicle with trees at the road side and no safety barrier.*

The second proposed example (see Fig.12) is a single vehicle traveling along a road segment of one km at a constant speed of 90 km/h, there is no roadside barrier and there are trees spaced every 5 meters on both sides of the road. For comparison, we can make the same assumptions as in example 1: diver failures are generated every second, driver distraction time is 5 seconds and 15 degrees deviations. Two crashes against trees, on both sides of the road, are generated by each one of the 40 considered positions of the vehicle. For the crashes at a speed of 90 km/h against the trees we have to consider the total kinetic energy of the vehicle (not only the $v_x$ component perpendicular to road direction):

$$\Delta K = \frac{1}{2}m(v)^2 = 312\ 500\ \text{(Joule)} \tag{17}$$

The total crash energy is: 25 000 000 (Joule), this energy value which is around 15 times that of example 1 and clearly shows how the proposed methodology would allow researchers to take into account also roadside barrier effects. The formula for total crash energy for example 2 is:

$$\text{Total crash Energy} = Z_{5\text{-}\theta\text{-}1/3} = \frac{L}{v \cdot \delta t} \cdot 2 \cdot \frac{1}{2}m(v)^2 * 1/3 \tag{18}$$

The deviation angle here again has practically no role (except for when the trajectory becomes so angulated that the vehicle could move through two trees without crashing).

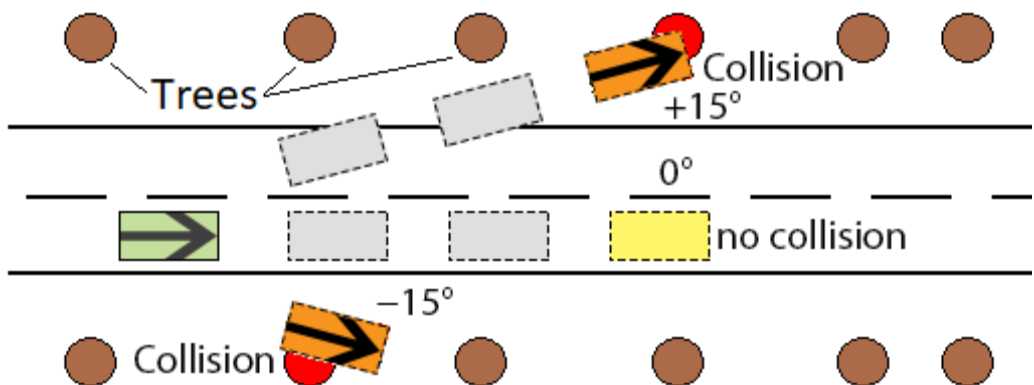

*Figure 12: Example 2.*

## 4. Microsimulation case studies

A general analysis of how the parameters affect results, in more complicated traffic scenarios, can only be done in microsimulation, given the practical difficulties of performing analytical calculation when more than one moving vehicle is present on a network. Three scenarios are presented in the following:

*4.1 Friction effects on safety in traffic flow theory. Is it true that significant speed differences do not affect traffic safety when a white (or yellow) line is drawn on the road pavement?*

The ancient Greeks were aware of causes and mitigation of friction. Fluid friction occurs between fluid layers that are moving relative to each other. This can cause viscosity and other friction forces in fluids. Friction can cause fluids to become unstable and can cause a transition to a turbulent flow regime. Traffic flow is the only case where it is possible to draw a white (or yellow) continuous line on the road pavement between two different flow speed regimes and supposedly avoid any friction and unwanted interactions. According to current surrogate safety indicators based on traffic conflicts a road design such as in Figure 13 is as safe as a road design with a central barrier since trajectories of opposing directions do not intersect. This is against common knowledge. Knapp *et al.*(2014) in the Road diet FHWA guide state that: "Four-lane undivided highways have a history of relatively high crash rates as traffic volumes increase".

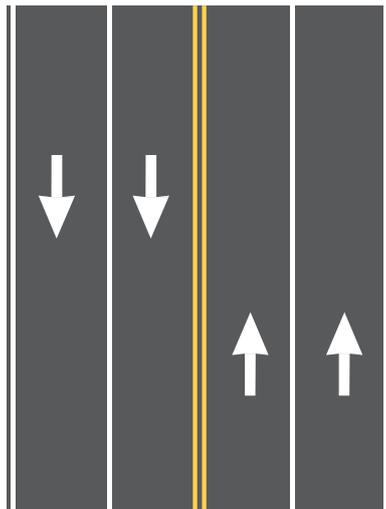

*Figure: 13 undivided four-lane highway.*

Many papers have been presented on the evaluation of surrogate safety measures based on microscopic or macroscopic traffic flow modeling that consider speed differential as a factor or as an indicator for crash risk (Kachroo and Sharma,2018). There is no published paper on surrogate safety indicators that can be applied to the speed differential between opposing traffic in undivided two (or four)-lane highways. In the seminal paper by Gettman and Head (2003) there is no mention of potential conflicts arising from this kind of traffic scenarios. In Laureshyn et al.(2010) it is introduced a classification of encounter states and the state where road users' paths do not overlap is considered to be safe unless the situation turns into a "collision course" possibly via a "crossing course" state. It is a fact that this transition from one state to another is not reproducible in current microsimulation unless driver error is introduced such as in the proposed approach. Moreover, it is debatable whether this weakness deserves a better investigation for all (numerous) applications that are based on real trajectories.

The conclusion is that the evaluation of undivided highway safety it is not possible in simulation with current surrogate safety measures based on conflicts and at the moment it has been explored only with inferential statistics based on crash data. Some examples are in: Council and Stewart (1999), Hauer at al. (2004); Dinu and Veeraragavan (2011), Park and Abdel-Aty (2015).

The following example will show how the proposed methodology can asses safety in these kinds of scenario. A simulation case study was created based on the situation depicted in Figure 14 relative to the road that runs alongside the sea in the city of Paola in south Italy.

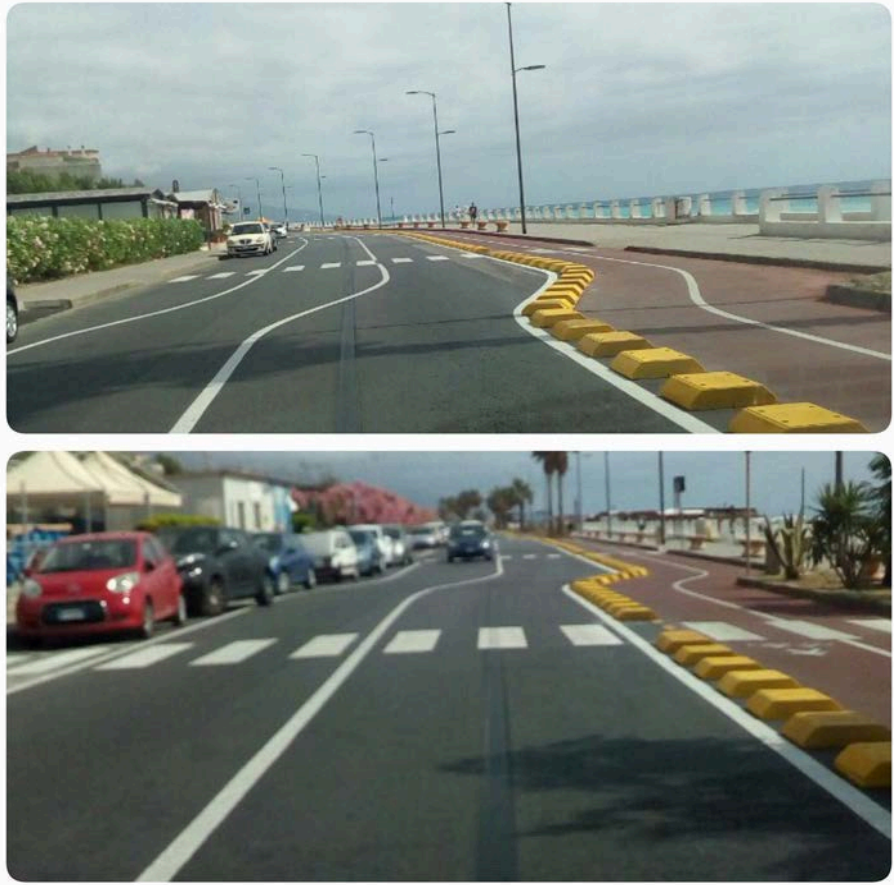

*Figure 14: Sea-side road in Paola, Italy 2018.*

A simulation was performed using VISSIM microsimulation package coupled with the SSAM package on the two networks depicted in Figures 15 and 16.

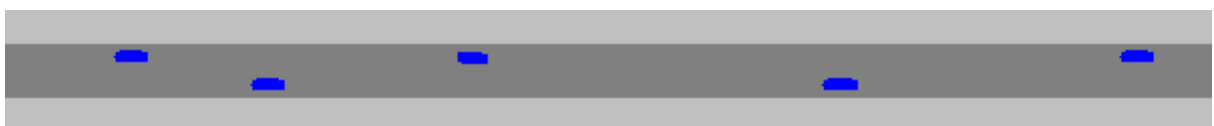

*Figure 15: Linear road.*

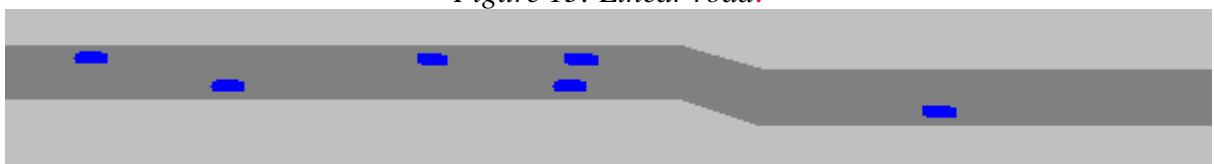

*Figure 16: Deviated Road.*

Simulated traffic conditions were the same in both scenarios: total length of the road 200 meters and traffic flow 500 veh./h. per direction. For both scenarios 20 simulation repetitions were carried out. Results of SSAM package with a TTC threshold of 1,5 sec. were equal for both scenarios: zero conflicts as can be seen in Figure 17.

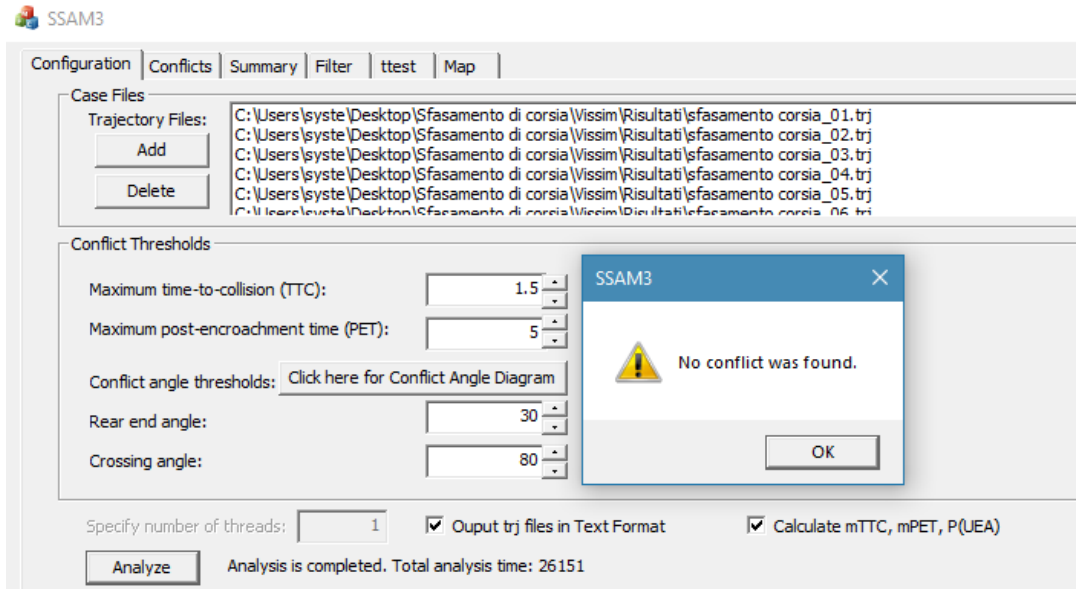

*Figure 17: SSAM Analysis on VISSIM simulation for first case study.*

The underling hypothesis in all current conflict techniques is that for a conflict between two vehicles to exist the trajectories must intersect. The case of Figures 14 and 16 is instead obviously dangerous since vehicles can potentially get involved in a high energy head on collision with traffic coming from the opposite direction. The proposed methodology instead, introducing driving errors in vehicles trajectories is able to evaluate the increased risk. Results, in fact, leads to a 30% worsening of safety, in terms of overall impact energy (Table T1), when the deviation is considered.

| Total collision energy [J] | | | | |
|---|---|---|---|---|
| Trajectory angle | Deviated road | Straight road | Difference | % Difference |
| Front | 378141 | 16336 | 361805 | **2214,8** |
| Left | 1948400 | 1770168 | 178231 | **10,1** |
| Right | 715 | 48 | 668 | **1394,9** |
| **Total** | 2327256 | 1786552 | 540704 | 1.1.1. 30,3 |

*Table 1: Total collision energy in VISSIM simulation.*

From the numerical results we can see how the incidents due to front and right trajectories have an important role in this proposed scenario since a distraction trajectory on the left of the travelling direction would bring vehicles away from a potential head on collision. In Figure 18 the collision points, in which the collisions start, are plotted and it is clearly possible to see an increase in point density near the deviated road section.

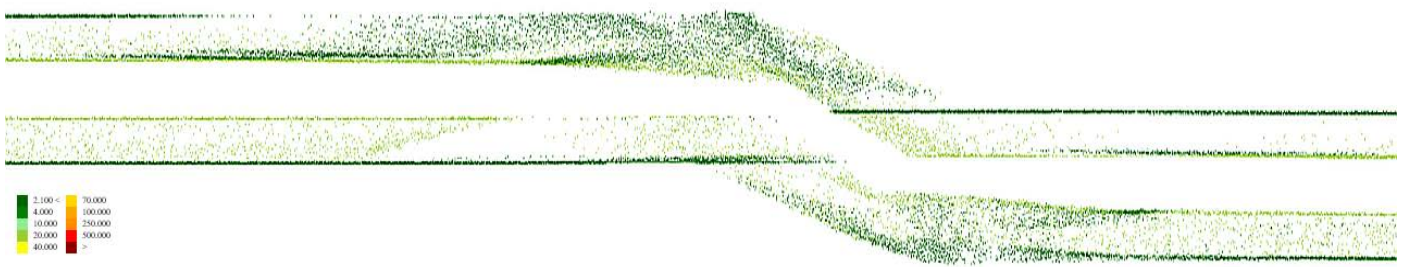

*Figure 18: Collision points.*

*4.2 Toll plaza layouts.*

The second simulated scenario concerns the analysis of two different toll plaza layouts. This scenario is based on the crash analysis carried out by Abuzwidaha and Abdel-Aty (2018) on different designs of hybrid toll plazas (HTP). In HTP express automated Open Road Tolling (ORT) lanes coexist with traditional toll collection lanes. The paper of Abuzwidah and Abdel-Aty (2018) compares the two different implemented solutions in Florida expressways: ORT deployed on the mainline and separate traditional toll collection to the side or traditional toll collection deployed on the mainline and separate ORT lanes to the side. The two main reported results are: that the first design (where ORT are deployed on the mainline) is safer and that the crash risk is approximately 23% higher in diverge areas than in merge areas.

An attempt to perform a risk estimation with our methodology has been carried on, replicating a similar scenario with the Tritone microsimulation package. Different total traffic flows values have been simulated on the networks depicted in Figure 19 and 20 considering 19% of the traffic on the traditional manual collection lanes.

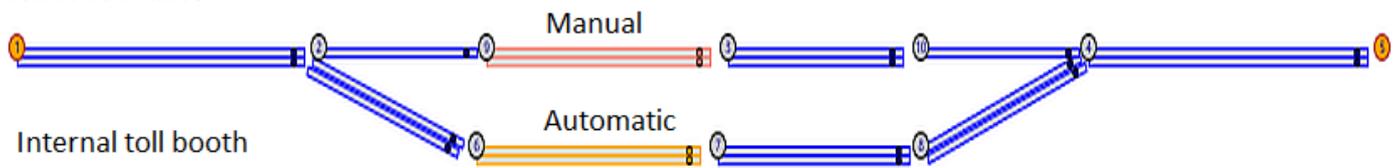

*Figure 19: First design (D1) of hybrid toll plaza layout (manual lanes on the mainline).*

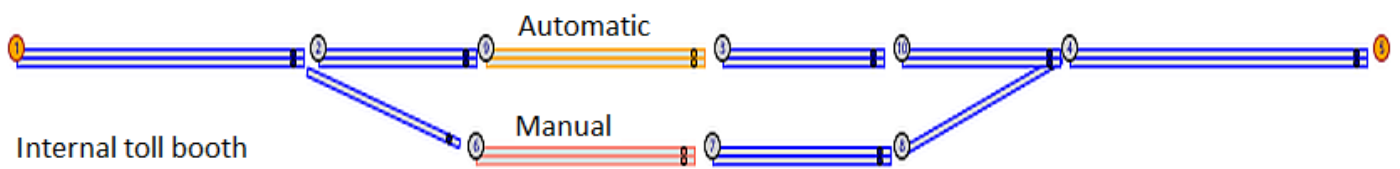

*Figure 20: Second design (D2) of hybrid toll plaza layout (automatic lanes on the mainline).*

Results in terms of total energy resulting from the application of our methodology, with a $Z_{3\text{-}15\text{-}1/3}$ indicator, are depicted in table 2 and, in accordance with traffic crash data of Abuzwidah and Abdel-Aty (2018), show an increased risk when the manual toll collection is deployed on the mainline.

| Traffic flow [v/h] | | | Total energy [J] | | | Safety |
|---|---|---|---|---|---|---|
| Total | Automatic toll collection | Manual toll collection | Manual collection on the mainline (D1) | Automatic collection on the mainline (D2) | Difference | % |
| 600 | 486 | 114 | 61304576 | 67775472 | 6470896 | 9,55 |
| 1200 | 972 | 228 | 218669984 | 258487232 | 39817248 | 15,40 |
| 1800 | 1458 | 342 | 1021998016 | 1160492288 | 138494272 | 11,93 |
| 2200 | 1782 | 418 | 1162934272 | 1332596992 | 169662720 | 12,73 |
| 2400 | 1944 | 456 | 1941774336 | 2412963072 | 471188736 | 19,53 |

*Table 2: Total collision energies in different flow scenarios.*

In Figures 21 and 22 it is possible to see the points in which the collisions take place and a visual (qualitative) confirmation of the second main finding of Abuzwidah and Abdel-Aty (2018) that the diverge areas are more critical as opposed to the merge areas.

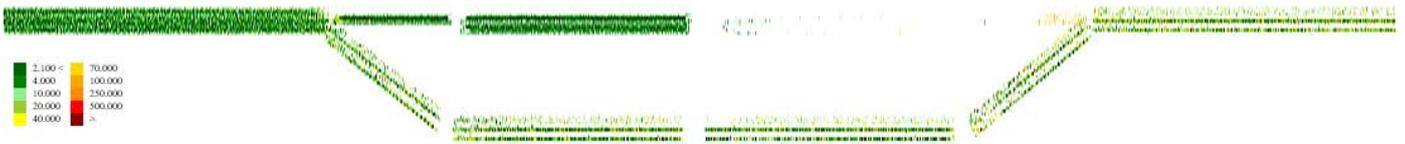

*Figure 21. Collision points in D1 design.*

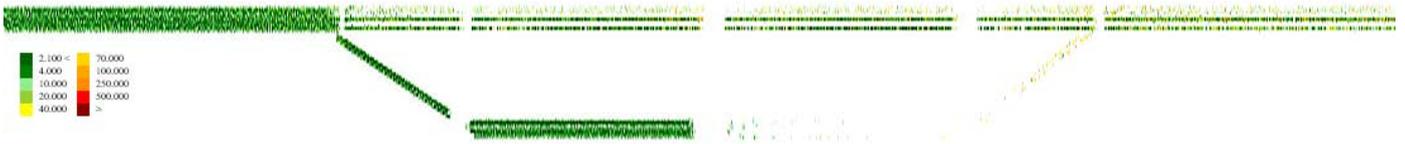

*Figure 22. Collision points in D2 design*

With the proposed methodology a map of the transportation network can be drawn by dividing the network into areas and by coloring each area according to the total energy of potential crashes that happen inside a given area. This visualization allows engineers to spot dangerous locations easily. In Figure 23 this diagram is presented for the two designs D1 and D2.

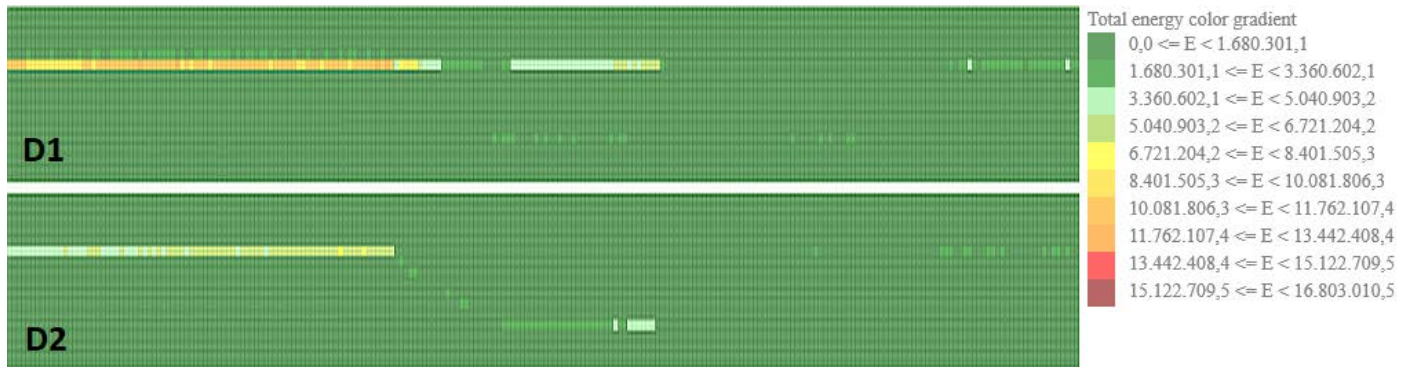

*Figure 23: Map of danger for D1 and D2 designs.*

Saad et al. (2018) present a study on the analysis of driving behavior at expressway toll plazas and it confirms previous results that the main problem at hybrid toll plazas is that drivers have to take the right lane and may perform sudden lane changing before the toll. The simulated scenario in our methodology, introducing driver errors, seems to be able to reveal these kinds of traffic flow turbulence and the connected risk.

Another example of the application of our methodology and the resulting danger map is in Figure 24 where an intersection is immediately spotted as a dangerous point on the traffic network.

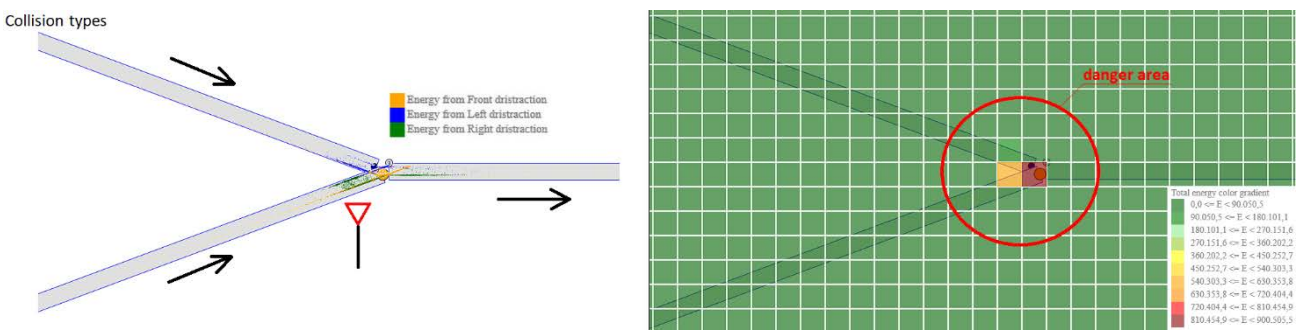

*Figure 24: Map of danger as potential crashes energy level.*

The small road network in Figure 24 has been simulated with 720 veh./h. on the main road and 360 veh./h. on the secondary road. It was modeled using the TRITONE software and elaborated using the methodology proposed in this article with a $Z_{3\text{-}15\text{-}1/3}$ indicator.

*4.3 three different intersection layouts*

Many papers have been presented for the evaluation of traffic safety at intersections with surrogate safety measures, among them: Stevanovic et al.,2013;Killi and Vedagiri,2014; Shahdah et al. 2015; Zhang et al., 2017. Some papers have been also presented to establish the safest design for given traffic volumes (Astarita *et al.* 2019) using simulation.

Following a similar approach, we introduce a three intersection layouts scenario, depicted in the following Figure 25. The simulations have been calibrated by sampling data from a single real intersection. The sampled intersection is located in an urban area and is an unsignalized intersection (case A). The intersection has also been studied altering the layout into a signalized intersection (case B) and a roundabout (case C) (keeping the same traffic flows).

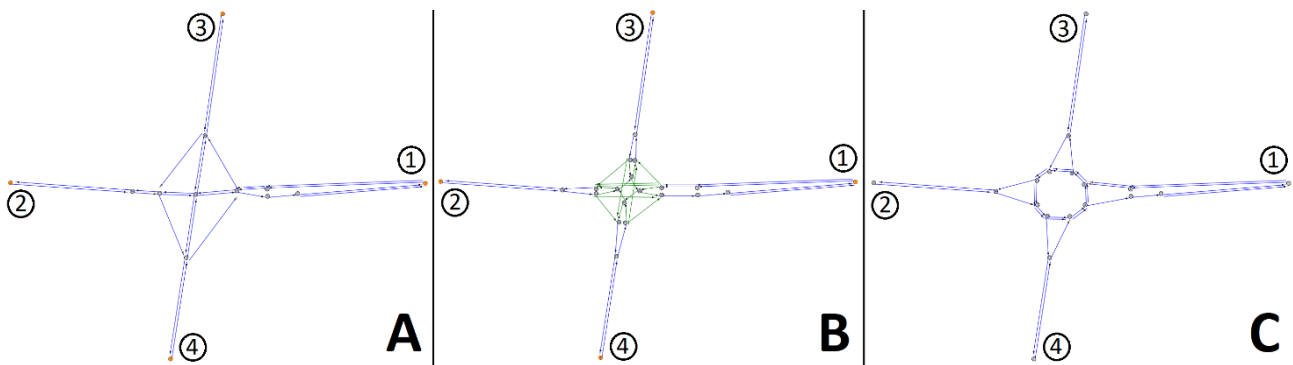
*Figure 25: Case studies*

The general expectation is that from case A to C, safety would increase in average conditions. The flows that were detected by manual counting and simulated are from 8.30 to 9.30 such as depicted in table I.

| Cars | | | | | Heavy Vehicles | | | | |
|---|---|---|---|---|---|---|---|---|---|
| O/D | 1 | 2 | 3 | 4 | O/D | 1 | 2 | 3 | 4 |
| 1 | 0 | 1073 | 183 | 124 | 1 | 0 | 48 | 14 | 13 |
| 2 | 465 | 0 | 74 | 133 | 2 | 13 | 0 | 3 | 0 |
| 3 | 76 | 75 | 0 | 95 | 3 | 5 | 8 | 0 | 3 |
| 4 | 121 | 215 | 148 | 0 | 4 | 8 | 8 | 7 | 0 |

*Table 3: O/D Matrices.*

From table I it is clear that in simulations also heavy vehicles were taken into account, in order to obtain a simulation closer to reality. The networks were modeled through the car-following models Wiedemann 99 (Wiedemann, 1991) and Gipps (1981) in both VISSIM and Tritone simulators. While the resulting trajectories were analyzed through a specific microsimulation add on software created ad hoc and based on the algorithm described in the previous sections.

*4.3.1 Results of simulations*

Simulations were carried out with different distraction time parameter: 3, 5, and 7 seconds and with a time step parameter of one second. This means that a simulated crash can happen at any instant between 1 second and the established distraction time.
The following two figures show results for Vissim and Tritone Microsimulation packages in terms of total energy of simulated crashes.
A higher energy can be associated with a greater risk of injuries or deaths. Absolute values of energy in the two different simulators are different (this reflects slightly different vehicle to vehicle interactions between different simulators), yet results in terms of relative values between different scenarios are very similar for the two simulators. Also results are not very sensible to a change in distraction time parameter (max distraction time in Figures 26 and 27). For all the selected distraction time parameters and for both simulators the intersections rank in this order of safety from the safest to the least safe: traffic light regulated, roundabout and unregulated intersection. In all scenarios, the unregulated intersection is the least safe while the traffic light is the safest. It must be noted that results are also a function of the given traffic flows. As can be seen in Figures 28 and 29, conditions of flow are very congested, in the traffic light scenario, and this increases safety since vehicles are always travelling at a very low speed.

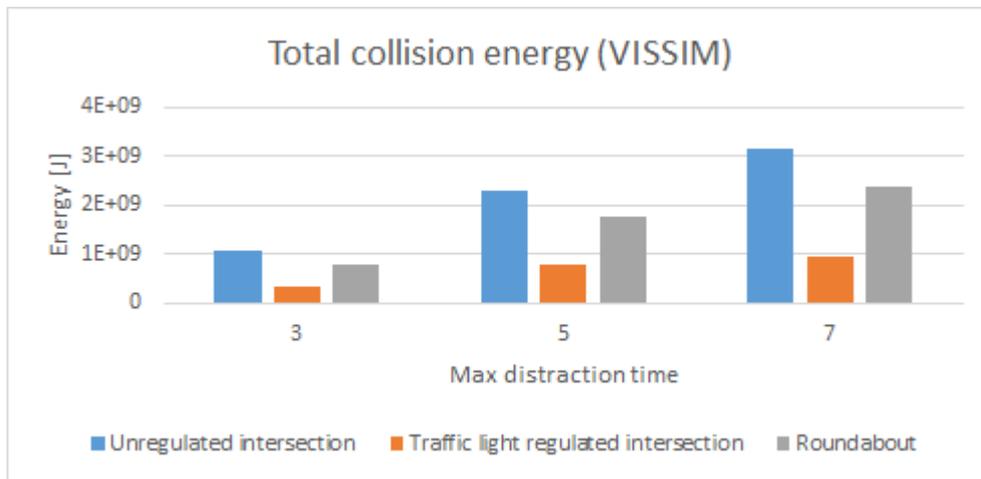

*Figure 26: Total collision energy in VISSIM for three scenarios and for different distraction times.*

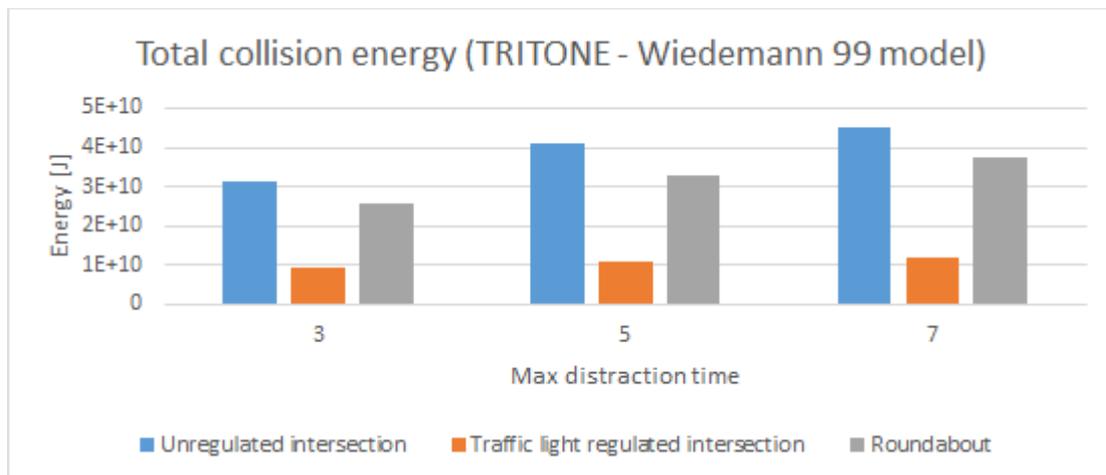

*Figure 27: Total collision energy in Tritone for three scenarios and for different distraction times.*

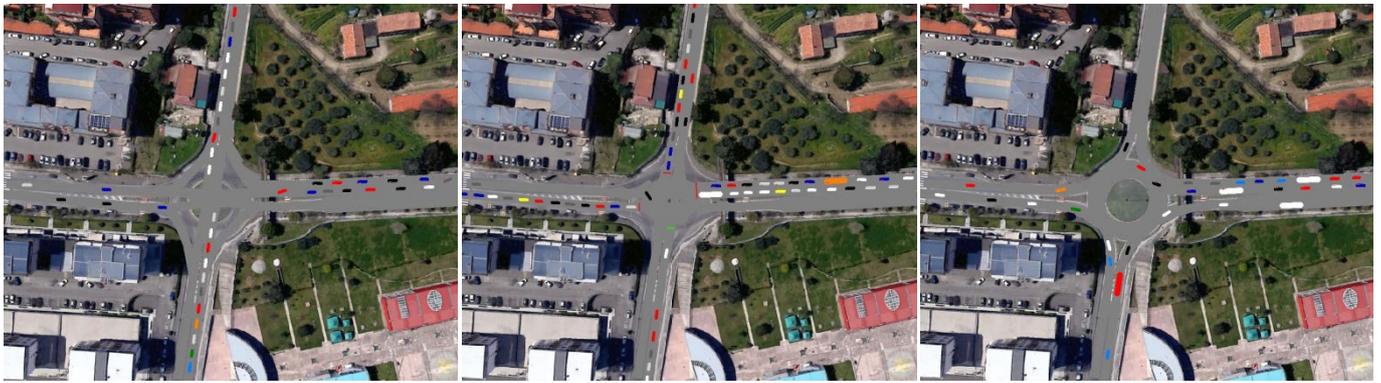
*Figure 28: The three simulated scenarios in VISSIM*

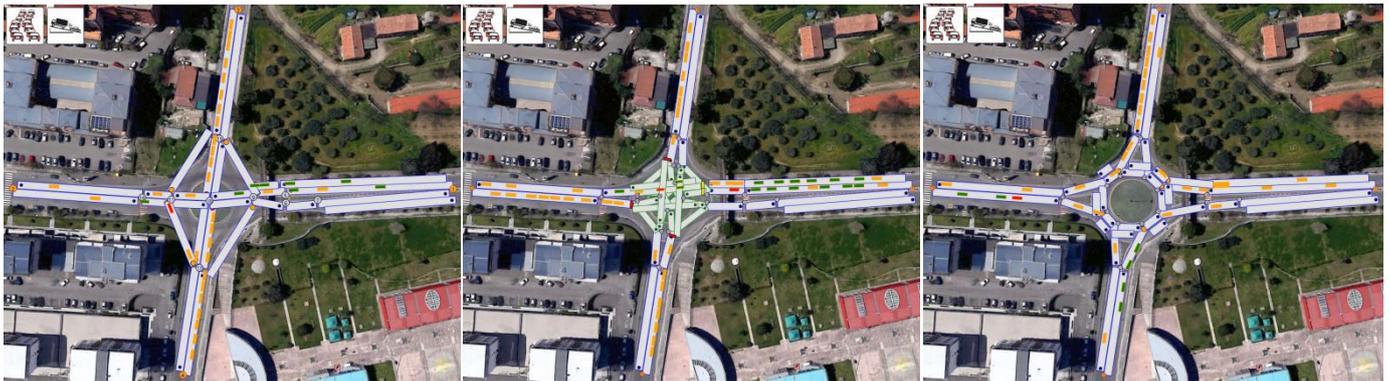
*Figure 29: The three simulated scenarios in TRITONE*

Results on the average time at which a crash occurs in the simulations are also similar between Tritone and VISSIM as can be seen in Figures 30 and 31:

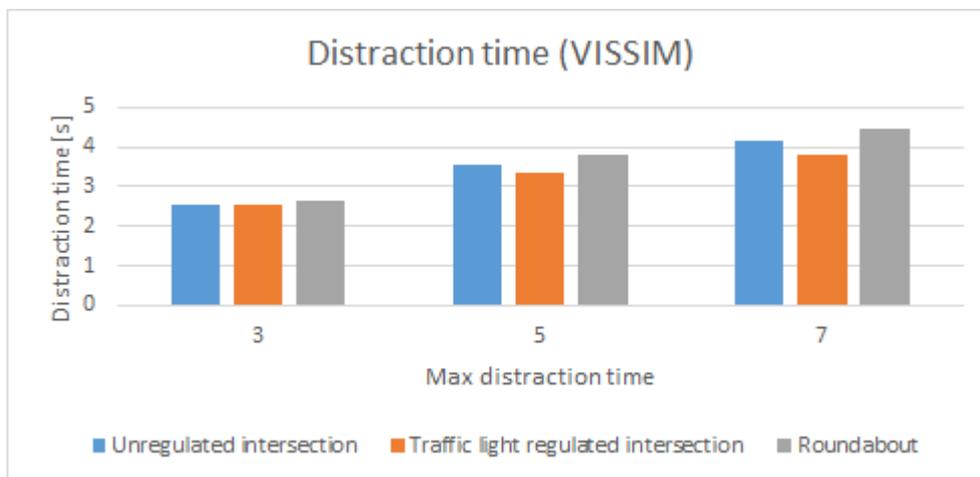
*Figure 30: Average distraction time before a crash occurs in VISSIM.*

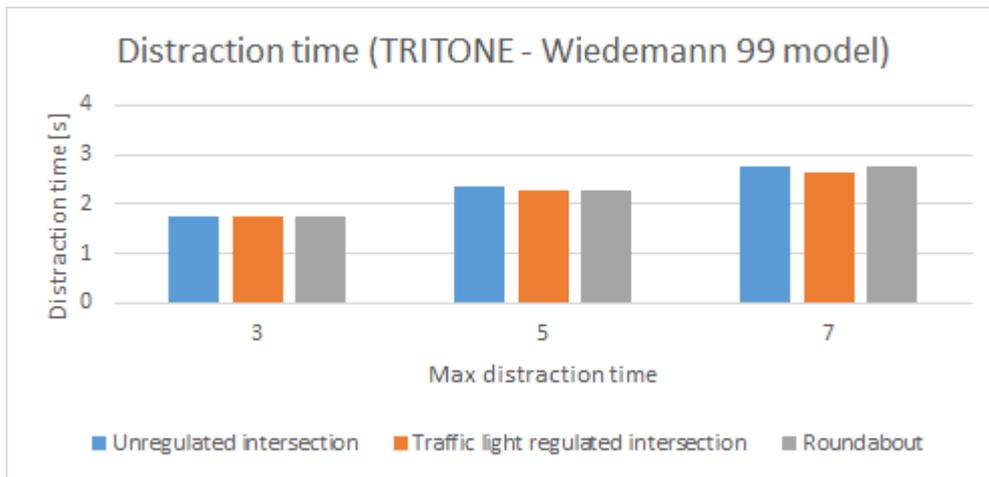

*Figure 31: Average distraction time before a crash occurs in Tritone.*

The total energy of simulated crash has been calculated, with a distraction time of 3 seconds, also by applying a threshold on the severity of crashes. In other words, only the crashes that have an energy value over threshold are considered and summed to calculate total energy. In Figure 32 and 33 are the total collision energies for VISSIM and Tritone:

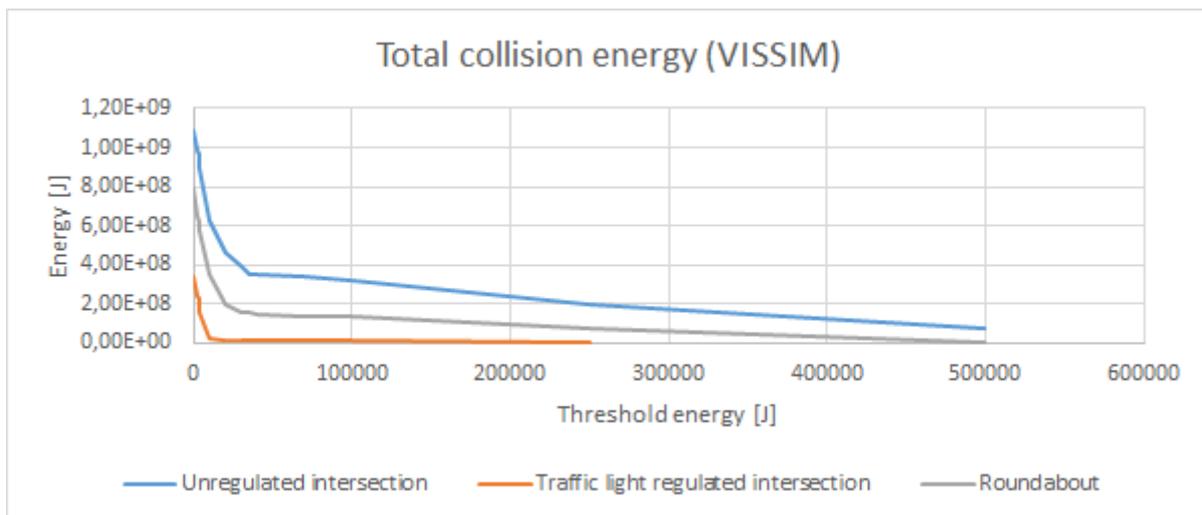

*Figure 32: Total collision energy in VISSIM removing all crashes below a given threshold.*

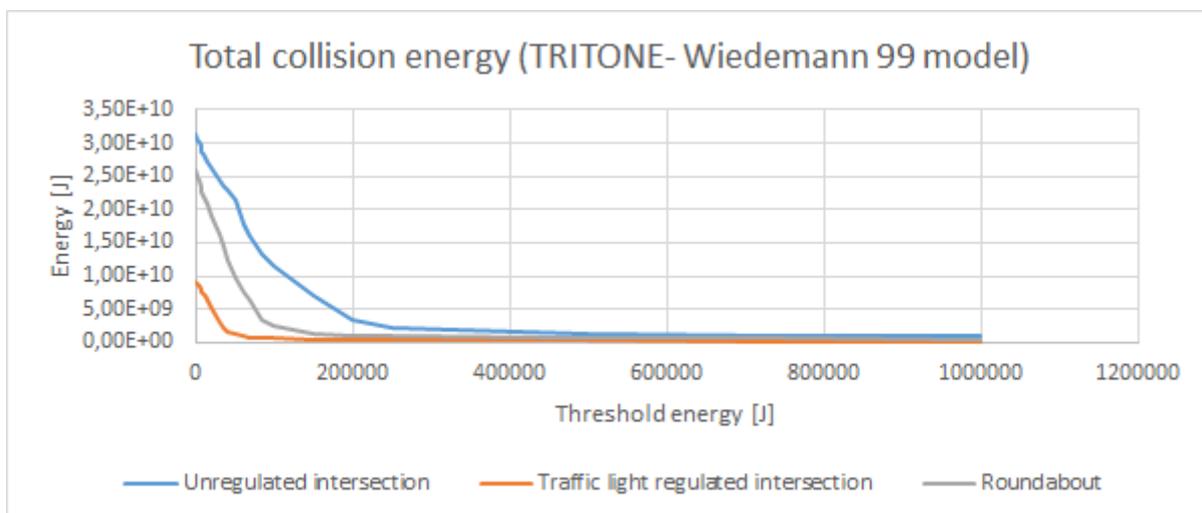

*Figure 33: Total collision energy in TRITONE removing all crashes below a given threshold.*

The graphs of Figures 32 and 33 show clearly how the ranking of the three scenarios are always the same in any case with both microsimulation packages.
When considering instead the average collision energy VISSIM and Tritone show different results as can be seen in the following Figures 34 and 35 depicting the average collision energy applying a threshold on the severity of crashes.

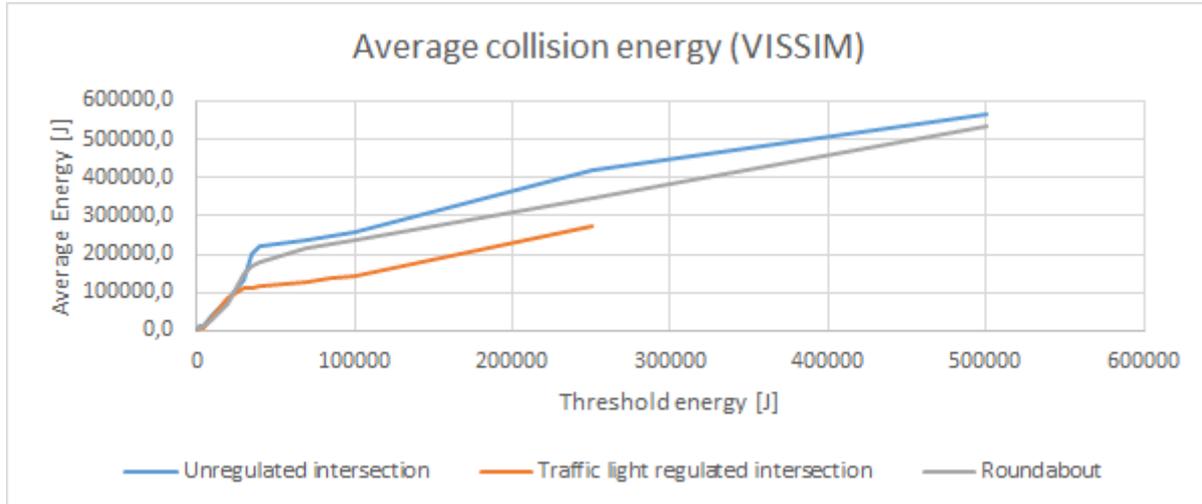

*Figure 34:Average collision energy in VISSIM removing all crashes below a given threshold.*

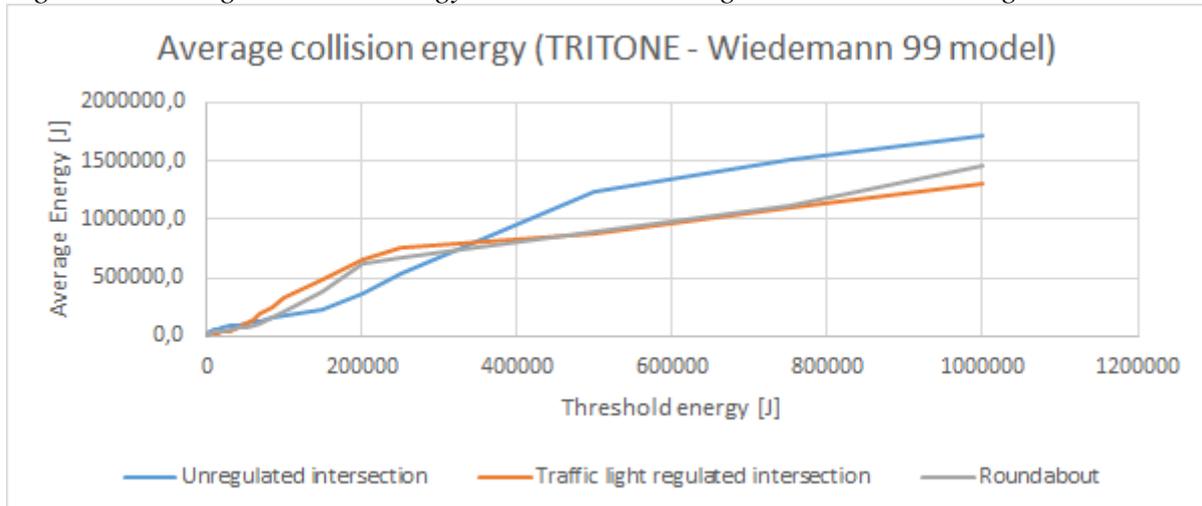

*Figure 35:Average collision energy in Tritone removing all crashes below a given threshold.*

### 4.3.2 Confrontation between the number of simulated crashes and SSAM conflicts

A confrontation was also performed between the proposed methodology and classic methodologies of estimation based on SSAM safety analysis. SSAM was applied to both Tritone and VISSIM in the three scenarios obtaining similar results in terms of number of conflicts. Conflicts were calculated on the basis of TTC indicator with a threshold of 1,5 sec.

The following Figure 36 shows again the same ranking in terms of safety levels for the three scenarios. In Figure 36 the percentage of the total number of conflicts (with classical methodology according to SSAM) and the total number of simulated crashes (among different scenarios) are shown for each scenario.

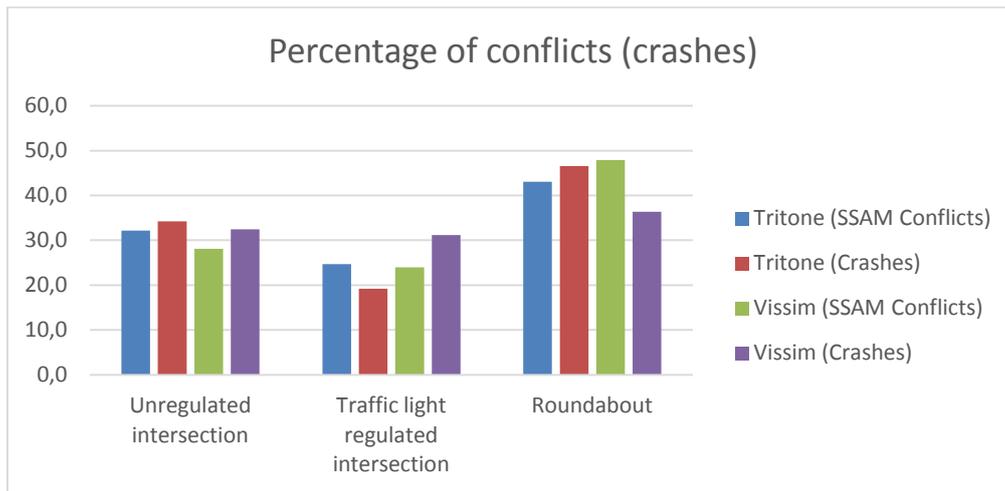

*Figure 36: Percentage of total conflicts(or crashes) for each intersection and for both microsimulation models.*

## 5 Discussion and conclusions

In this paper, a general methodology for introducing the driver error in simulation and for evaluating resulting potential crashes is presented. The proposed methodology is quite primitive in both the simulation of driver errors and in crash dynamics simulation, yet it presents very innovative features introducing the possibility of simulating driving errors in a microscopic environment.

The simplicity of the proposed approach makes it easily reproducible and applicable on top of any microsimulation package that can produce trajectories of vehicles.

Moreover, the proposed approach generates surrogate safety indicators that can be applied legitimately on both simulated and real trajectories as alternative surrogate safety measures and compared with existing indicators that are not able to consider conflicts between vehicles moving on nonconflicting trajectories and with roadside obstacles and barriers.

The results of this paper indicate that the safety level of very common traffic scenarios, which cannot be assessed with traditional surrogate safety measures, based on conflict indicators, can instead be easily captured with the proposed methodology.

The proposed indicators are, in fact, more comprehensive than traditionally used conflict indicators since they are able to capture a wider range of conflicts that those commonly evaluated in traditional indicators.

The assessment of roadside barriers safety becomes possible in the proposed framework with common microsimulation packages.

A more detailed analysis of potential crash consequences is also possible by exploiting the richness of crash dynamics simulation.

A different and new insight into an important issue such as traffic safety evaluation with microscopic simulation becomes possible when the introduction of driver errors (or sudden road anomalies) allows road crashes in a microscopic setting to be simulated.

Common simulation techniques do not allow representation of a wide range of erroneous driver behaviors that can lead to a crash. Even common shunting crashes that are caused by any internal or external cause that can force a leading driver to stop the car, with an abrupt extreme braking maneuver, are not properly represented with common simulation techniques.

More common erroneous driver behaviors that can be simulated by the proposed approach are:
-Drivers speeding at a red light or in general approaching an intersection without taking into proper consideration other incoming vehicles.
-Drivers taking the wrong lane, driving on the wrong side of the road or (worst) in one way streets against the correct traffic direction.
-Drivers driving off road in isolated accidents against external obstacles or traffic barriers.
-Accidents where a vehicle hits first a barrier or an obstacle and then bouncing back on the road hitting or getting hit by another car.
-Accidents that involve more than one vehicle and the connected real crash dynamic.
-Accidents caused by overtaking maneuvers where there is not enough time leeway.

The proposed first implementations of the methodology in the presented case studies show accordance with empirical expectations, with crash data and with commonly used conflicts indicators techniques (in the situations where conflict indicators can be applied).

Crash severity is finally taken into account explicitly shedding some light on the real consequences and on the safety of traffic scenarios that it was not possible to simulate properly.

The total impact energy proposed in the case studies is only one of the possible surrogate measures of risk that it is possible to calculate once driver errors are introduced. By simulating (or knowing) the vehicle mass distribution it would be an easy step to calculate the exact Delta-V of every potential crash bringing an estimate of the casualties or injuries using (11).

Moreover, the methodology presented can be easily applied to evaluate the safety impact of autonomous vehicles in a mixed environment. Autonomous vehicles of the future will be able to drive without committing human errors yet they will have to face the possibility of human errors. A safety assessment using microsimulation of different mixed road scenarios would be realistic only properly introducing human error as the main differentiating factor between human driven and autonomous vehicles.

The presented methodology will allow important safety and economic evaluations of new mixed traffic interventions and policies such as: new traffic rules in a mixed traffic reality, roadside barriers design, new connected traffic signals and many other innovative systems that will have to deal with a mixed circulation of old vehicles and connected and autonomous vehicles.

This evaluation, of course, would have to be scaled relative to the simulated scenario as a function of the traffic flow distribution over a period of time. The indicators proposed in the case studies examine a driving error rate of one error per second for each driver. In a real scenario, the driving error rate for second is a very low number though the number of seconds travelled by all drivers are very high. A scaling of the proposed procedure on experimental data and a calibration of the indicators parameters its outside the scope of this first methodological paper.

The implementations of the proposed methodology in the proposed case studies should be only considered as illustrative examples of the potentiality of introducing driver errors into a microsimulation environment so that the concepts presented in this seminal paper may stimulate new questions and give new insights that hopefully will lead to further investigations and better infrastructure designs.

# 6 Acknowledgements

We wish to thank Giuseppe Guido and Alessandro Vitale for the useful discussions, and Giovanna Imbrogno for suggesting the Paola sea-side road case.